\documentclass[reprint,superscriptaddress,pra,nofootinbib,preprintnumbers]{revtex4-2}

\usepackage[utf8]{inputenc}
\usepackage{amsmath}
\usepackage{xcolor}
\usepackage{graphicx}
\usepackage{dsfont}
\usepackage{physics}
\usepackage{placeins}
\usepackage{amssymb}
\usepackage{mathtools}
\usepackage{bbm}
\usepackage[caption=false]{subfig}
\usepackage[colorlinks=true, urlcolor=blue,citecolor=blue,anchorcolor=blue]{hyperref}

\providecommand{\CO}{\mathcal{O}}

\providecommand{\CW}{\mathcal{W}}

\providecommand{\CQ}{\mathcal{Q}}
\providecommand{\CP}{\mathcal{P}}

\providecommand{\uone}{${\rm U}(1)$\,}

\newcommand{\ie}{{\textit{i.e.\,}}}

\begin{document}
\title{
Overcoming exponential scaling with system size in Trotter-Suzuki \\ implementations of constrained Hamiltonians: 2+1 U(1) lattice gauge theories\\
}
\author{Dorota M. Grabowska}
\email{dorota.grabowska@cern.ch}
\affiliation{Theoretical Physics Department, CERN, 1211 Geneva 23, Switzerland}

\author{Christopher Kane}
\email{cfkane@arizona.edu}
\affiliation{Department of Physics, University of Arizona, Tucson, AZ 85719, USA}

\author{Benjamin Nachman}
\email{bpnachman@lbl.gov}
\affiliation{Physics Division, Lawrence Berkeley National Laboratory, Berkeley, CA 94720, USA}

\author{Christian W. Bauer}
\email{cwbauer@lbl.gov}
\affiliation{Physics Division, Lawrence Berkeley National Laboratory, Berkeley, CA 94720, USA}

\preprint{CERN-TH-2022-133}
\date{\today}

\begin{abstract}
For many quantum systems of interest, the classical computational cost of simulating their time evolution scales exponentially in the system size. 
At the same time, quantum computers have been shown to allow for simulations of some of these systems using resources that scale polynomially with the system size.
Given the potential for using quantum computers for simulations that are not feasible using classical devices, it is paramount that one studies the scaling of quantum algorithms carefully.
This work identifies a term in the Hamiltonian of a class of constrained systems that naively requires quantum resources that scale exponentially in the system size.
An important example is a compact \uone gauge theory on lattices with periodic boundary conditions. 
Imposing  the magnetic Gauss' law \textit{a priori} introduces a constraint into that Hamiltonian that naively results in an exponentially deep circuit.
A method is then developed that reduces this scaling to polynomial in the system size, using a redefinition of the operator basis.
An explicit construction of the matrices defining the change of operator basis,  as well as the scaling of the associated computational cost, is given.

\end{abstract}

\maketitle

\section{Introduction}

Constrained Hamiltonians are ubiquitous in fundamental physics, as any gauge theory -- general relativity, electromagnetism, Yang–Mills, string theory -- all result in such Hamiltonians. 
Unfortunately, constrained Hamiltonians pose a number of difficulties, both theoretical and practical. 
On the theoretical side, it is not always clear how to quantize the theory such that all constraints are properly implemented in a self-consistent manner \cite{Dirac1964, Henneaux1992, bojowald_2010}. 
On the practical side, as will be discussed in this paper, constrained Hamiltonians can result in a large degree of coupling, resulting in resource-expensive simulations.

This work will mostly be concerned with the ``degree of coupling'' (DoC) between various terms in the Hamiltonian. The DoC of a given term is defined as the maximum number of operators that need to be included in a single term of a Suzuki-Trotter represenation~\cite{Suzuki:1976be,10.1143/PTP.56.1454,10.2307/2033649}. Furthermore, the DoC of a Hamiltonian is defined as the maximum DoC of any given term within the Hamiltonian itself.
For example, for a Hamiltonian of the form $H = \sum^{N_\CO}_{ij} \CO_i \CO_j$, where the operators $\CO_i$ are  local operators and $N_\CO$ is the total number of such operators, the ${\rm DoC} = 2$, independent of the total number of operators. %
On the other hand, a Hamiltonian of the form $H = F(\sum_i^{N_{\CO}} \CO_i)$ will have 
\begin{align}
    {\rm DoC}[F; N_\CO] \equiv \min \left[N_\CO, \deg(F)\right]
    \label{eq:DoCH}
\end{align}
where $\deg(F)$ is the polynomial degree of the function $F$; note that a non-polynomial function has $\deg(F) = \infty$. Throughout this work, the arguments of a given DoC will sometimes be suppressed for brevity.

As we will discuss, constraints often give rise to Hamiltonians where the DoC scales with the system size, which in turn naively creates an exponential scaling in the gates required to simulate the time evolution of such a system.
Note that the degree of coupling is different from the non-locality of the system, which typically has to do with the minimum distance between two entangled qubits.
A well-known, and phenomenologically relevant, class of constrained Hamiltonians are gauge theories formulated on space-time lattices with periodic boundary conditions.
In such theories, the Hilbert space is divided into different charge sectors, with the gauge invariant Hamiltonian not including any interactions that migrate between the gauge sectors.
However, due to noisy machines as well as a practical inability to exactly simulate time evolution, one needs to contend with the different charge sectors. 
Formulating these theories using only the physical subspace introduces the constraints with which this paper is concerned.
It is important to note, however, that there are proposals for quantum simulations that do not rely on constructing Hamiltonians that span only the physical Hilbert space. 
The basic idea of these methods is to introduce an energy penalty term in the Hamiltonian that suppresses transitions that are gauge variant~\cite{PhysRevLett.107.275301,PhysRevLett.109.125302,Hauke2013}. For more details on this approach see Refs.~\cite{Halimeh:2019svu, Lamm:2019bik, Lamm:2020jwv, Tran:2020azk, PhysRevLett.107.275301, Banerjee:2012pg, Kasper:2020owz, PhysRevX.3.041018,  Kuhn2014, Stannigel:2013zka}.

As a test case, the dual-basis formulation of a \uone lattice gauge theory in two spatial dimensions\footnote{For the remainder of this manuscript, we will use `D+1' dimensions to refer to a theory that has D spatial dimensions and one time dimension.} is studied \cite{Drell:1978hr,Kaplan:2018vnj, Haase:2020kaj, Bender:2020ztu}. While this theory does not appear in nature, it is a very useful case study. In particular, the compact formulation is expected to display linear confinement, similar to QCD in three spatial dimensions \cite{PhysRevD.10.2445, PhysRevD.11.395, PhysRevD.13.1043, Polyakov:1975rs, Polyakov:1976fu, Drell:1978hr}. As discussed in Ref.~\cite{Kaplan:2018vnj}, for such a theory, the set of magnetic fields on each plaquette of the lattice is generally over-complete, since the magnetic Gauss' law constrains the total magnetic flux through any closed surface to vanish.
For a 2+1 dimensional lattice with periodic boundary conditions, this leads to one constraint. For open boundary conditions, there is no closed surface and so no constraint appears. 
In 3+1 dimensions, the number of constraints for periodic boundary conditions is $2 + N$, where $N$ is defined as the total number of lattice sites; in this dimension, it is also the total number of cubes in the volume~\cite{Kaplan:2018vnj}. 
The number of constraints can be understood by realizing that there are $N-1$ independent constraints on single-cell cubes and then three extensive constraints involving a quadratic number of plaquette operators, one for each plaquette orientation (in the $\hat x-\hat y$ plane, $\hat x-\hat z$ plane and $\hat y-\hat z$ plane); note that these three extensive constraints do not appear for open boundary conditions. 
Solving the constraints and writing one of the operators in a given closed surface in terms of the other operators in that same surface can remedy the redundancy of the basis. 
Doing so, however, introduces terms in the Hamiltonian that couple together many operators at different lattice sites, naively giving rise to exponentially scaling circuits 
on a quantum device.

This manuscript presents a general method for breaking this exponential scaling in system size, based on the idea of reducing the degree of coupling via a judiciously chosen change in the operator basis. 
This paper is organized as follows: Section~\ref{sec:CoB} presents a general method for carrying out an operator redefinition that provides an exponential reduction in the degree of coupling of the system. 
In Section~\ref{sec:Uone} this method is applied to the case of a compact \uone lattice gauge theory with periodic boundary conditions, where a global constraint arises from the magnetic Gauss' law. 
The conclusions are presented in Section~\ref{sec:concl}.  

\section{Exponential reduction in the degree of coupling}
\label{sec:CoB}
It is instructive to start with a general Hamiltonian that has a maximal degree of coupling, in order to see how the exponential scaling arises. 
This Hamiltonian is given by 
\begin{align}
H &= \sum_{i=1}^N\left( f \left[\CQ_i\right]+g \left[\CP_i\right]\right) + F\left[\sum_{i= 1}^N \CQ_i \right] +G\left[\sum_{i= 1}^N \CP_i \right] \nonumber \\
&\equiv H_\CQ[\CQ_1,...,\CQ_N] + H_{\CP}[\CP_1,...,\CP_N]\,,
\label{eq: ArbitraryHam}
\end{align}
where $f, g, F,$ and $G$ are arbitrary functions; note that this manuscript will always use square brackets to denote arguments of these functions, for ease of  legibility. Additionally, $\CQ_i$ and $\CP_i$ are operators acting on single lattice sites, and $N$ is the system size; 
In lattice field theories, $N$ would be the lattice volume.

As will be explained shortly, the naive computational cost of implementing a given Hamiltonian scales exponentially with its DoC.
The DoC for the first two terms, which only depend on a single local operator $\CO_i \in \{\CQ_1,...,\CQ_N, \CP_1,...,\CP_N\}$, is
\begin{align}
    {\rm DoC}[f; 1] = {\rm DoC}[g; 1] = 1 \, ,
\end{align}  
where we use the notation set in Eq.~\eqref{eq:DoCH} that $\text{DoC}[X;n]$ is the Degree of Coupling for the function $X$, which depends on $n$ operators. This is not true for the last two terms, where the DoC depends on the number of operators in the system $N$ and whether either $F$ or $G$ are non-polynomial.
In particular, the DoC for the third and fourth term is given by 
\begin{align}
    {\rm DoC}[F; N] &= \min \left[N, \deg(F)\right], \nonumber
    \\ 
    {\rm DoC}[G; N] &= \min \left[N, \deg(G)\right] \, ,
\end{align} 
respectively. As it is the DoC of the total Hamiltonian that determines the complexity of implementation, the relevant DoC is given by the maximum of these two and so 
\begin{align}
    {\rm DoC} = \min \left[N, \max\left[ \deg(F), \deg(G)\right]\right]\, .
\end{align} 
Therefore, as long as either $F$ or $G$ is non-polynomial, the DoC of the total Hamiltonian is $N$ and, as will be shown, the computational cost of implementing this Hamiltonian will scale exponentially with the system size $N$. However, this analysis also points towards a method for reducing this scaling. Namely, there is a maximal imbalance between the DoC of the different terms in the Hamiltonian as it is currently written. However, a change of basis that better balances the DoC  between the terms with ${{\rm DoC}=1}$ and ${{\rm DoC}=N}$ such that the maximum DoC is reduced would also reduce the overall cost of implementing $H$. The main focus of this work is to present a particular operator basis change such that the DoC of any given term in the Hamiltonian is $\mathcal{O}(\log_2 N)$, which reduces the computational cost of implementing $H$ from exponential in $N$ to polynomial in $N$.

The naive computational cost of implementing this Hamiltonian can be estimated by specifying the properties of the operators $\CQ_i$ and $\CP_i$. In particular, it is assumed that $\CQ$ and $\CP$ are conjugate variables, such that they obey the commutation relations 
\begin{align}
\label{eq:CommmuteAssu}
\left[\CQ_i, \CP_j \right] = i\, \delta_{ij}
\,.
\end{align}
This implies that the basis  $\ket{\CQ_i}$ and the basis $\ket{\CP_i}$ are related by Fourier transform and that operators on different sites commute.
Therefore, one method for simulating the Trotterized time-evolution of this Hamiltonian on a quantum computer is to implement the matrices
\begin{align}
\braket{\CQ_i|e^{i H_\CQ}} {\CQ_j} \,\, \text{and} \,\, \braket{\CP_i|e^{i H_\CP}} {\CP_j}\,,
\end{align}
and use a (fast) Fourier Transform to switch between the two bases. 

Assuming that the each operator $\CQ_i$ and $\CP_j$ acts on a Hilbert space spanned by $n_q$ qubits (at a given lattice site), the dimension of the spanned Hilbert space is of size $2^{n_q}$. 
It is known that the number of gates required to implement an arbitrary diagonal matrix of dimension $2^{n_q}$ is exponential in $n_q$; for example, without using ancillary qubits, the count is $2^{n_q+1}-3$ gates~\cite{Bullock2003}. 
Therefore, exponentiation of each of the terms with ${\rm DoC} = 1$ in Eq.~\eqref{eq: ArbitraryHam} requires $\CO(2^{n_q})$ gates.
Implementing all of the terms with ${\rm DoC} = 1$ requires $\CO(N 2^{n_q})$ gates; however, as
these gates can be run in parallel due to the assumption in Eq.~\eqref{eq:CommmuteAssu}, the depth of the circuit is therefore $\CO(2^{n_q})$. 

Unfortunately, this is no longer true for the terms with maximal DoC, as they tie together all $N$ operators and thus require all $n_q N $ qubits. 
Therefore, implementing the term with maximal DoC requires $\CO(2^{n_q N})$ gates. 
This is the origin of the exponential scaling with $N$ that makes implementing this constrained Hamiltonian so costly. 

\subsection{General reduction method}
\label{ssec:GenCoB}
The exponential scaling in system size, resulting from a term that is a function of all the operators, can be broken by a judiciously chosen change of operator basis. 
One can define a new set of operators, via
\begin{align}
\CQ'_i = \CW_{ij} \CQ_{j} \qquad 
\CP'_i = \CW_{ij} \CP_{j} 
\,,
\label{eq:CoB}
\end{align}
where $\CW$ is an orthogonal matrix. These new set of operators obey the same commutation relations
\begin{align}
\left[\CQ'_i, \CP'_j \right] &= \CW_{ik}\CW_{jl}[\CQ_k, \CP_l] \nonumber \\
&= i \CW_{ik}\CW_{jl} \delta_{kl} \nonumber \\
&= i \CW_{ik}\CW_{jl} \nonumber \\
&= i \delta_{ij}
\,,
\end{align}
where the last line is a consequence of the orthogonality of the $\CW$. 
This implies that the eigenbases of $\CQ'_i$ and $\CP'_i$ are still related by a Fourier transform.
If the matrix $\CW$ is chosen such that not all of the operators appear within the same function $F$ or $G$, then the DoC of these terms will be reduced. 
However, this will also increase the DoC of the terms involving the functions $f$ and $g$. 
Therefore, $\CW$ has to be judiciously chosen to balance these two effects. 
Roughly speaking, in order to break the exponential scaling, the maximum DoC for any one term must scale logarithmically in $N$ (or as a polynomial of $\log_2 N$).
The remaining discussion will demonstrate how this can be achieved. In particular, in this subsection and the following, a specific form for $\CW$ will be proposed that will then be proven to provide this necessary tuning.

Consider the orthogonal matrix $\CW$ to be block diagonal in structure
\begin{align}
    \CW = \begin{pmatrix}
            \Omega^{(1)} & 0 & 0 & \dots &0 \\
            0& \Omega^{(2)} & 0 & \dots& 0  \\
            \vdots& \dots & \ddots & \vdots & \vdots  \\
            0& 0 & 0& \dots& \Omega^{(N_{S})}   \\
\end{pmatrix}
\,,
\end{align}
where $N_S$ is the number of sub-blocks and $\Omega^{(i)}$ are orthogonal matrices of dimension $d_{(i)}$. 
With this, the Hamiltonian $H_\CQ$ in the rotated basis (indicated by the superscript ``rot'') can be written as
\begin{align}
H^{\text{rot}}_\CQ =&  \sum_{i = 1}^{N}f \left[\sum_{j = 1}^{N}\CW_{ij} \CQ_j\right]  + F\left[ \sum_{i = 1}^{N}\sum_{j = 1}^{N} \CW_{ij} \CQ_j\right] \nonumber \\
=&\sum_{i = 1}^{N_S}\sum_{k = 1}^{d_{(i)}}f\left[ \sum_{j = 1}^{d_{(i)}} \Omega^{(i)}_{kj} \CQ_{D_{(i)}+j-1}\right]
\nonumber\\
&+ F\left[ \sum_{i = 1}^{N_s}\sum_{k, j = 1}^{d_{(i)}} \Omega^{(i)}_{kj} \CQ_{D_{(i)}+j-1}\right]
\,,
\label{eq:HQRot}
\end{align}
where $D_{(i)}$ labels the row and/or column in the matrix $\CW$ where a sub-block $\Omega^{(i)}$ begins,
\begin{align}
D_{(i)} = 1+ \sum_{n = 1}^{i-1}d_{(n)} \, ,
\label{eq:DDef}
\end{align}
and the summation over repeated indices has been made explicit for clarity. 
The Hamiltonian $H_\CP^\text{rot}$ will take on a corresponding form. 
In order to reduce the maximal DoC, the rotation matrix needs to reduce the number of terms within the function $F$. 
This can be achieved by choosing all elements of the first column of each $\Omega^{(i)}$ to be identical; the veracity of this choice can be seen in the following way.
Since all columns of orthogonal matrices are vectors of unit length, if all the elements of each $\Omega^{(i)}$ are identical, this implies $\Omega^{(i)}_{k1} = 1/\sqrt{d_{(i)}}$. 
Additionally, orthogonality implies the dot product of any two different columns is zero. Because the first column contains only identical entries, this implies the sum of all other individual columns must be zero, \textit{i.e.}
\begin{align}
\sum_{i=1}^{d_{(k)}} \Omega^{(k)}_{ij} = \sqrt{d_{(k)}} \, \delta_{1j} \, .
\end{align}
With this choice, the argument of $F$ becomes
\begin{align}
\sum_{i = 1}^{N_s}\sum_{k, j = 1}^{d_{(i)}} \Omega^{(i)}_{kj} \CQ_{D_{(i)}+j-1} &=\sum_{i = 1}^{N_s}\sum_{j = 1}^{d_{(i)}} \sqrt{d_{(i)}}\delta_{1j} \CQ_{D_{(i)}+j-1} \nonumber \\
&=\sum_{i = 1}^{N_s} \sqrt{d_{(i)}} \CQ_{D_{(i)}}
\,,
\end{align} and so the term that had the maximal DoC becomes
\begin{align}
F\left[\sum_{i = 1}^N \CQ_i\right] \Rightarrow F\left[\sum_{i = 1}^{N_S}\sqrt{d_{(i)}}\, \CQ_ { D_{(i)}}\right]
\,.
\end{align}
This implies that the DoC is no longer $N$ but $N_S$ and therefore, the number of gates needed to implement this term is reduced to $\CO(2^{n_q N_S})$. 
However, the exponential volume scaling will persist unless $N_S \lesssim (\log_2 N)^p$, for some power $p$.

While the change of basis has decreased the DoC for the function $F$, it has done so at the cost of increasing the DoC for the function $f$. 
In particular, the $i^\text{th}$ term with $\rm {DoC} = 1$ becomes
\begin{align}
f\left[\CQ_i\right]\Rightarrow  \sum_{k = 1}^{d_{(i)}}f\left[ \sum_{j = 1}^{d_{(i)}} \Omega^{(i)}_{kj} \CQ_{D_{(i)}+j-1}\right]
\,,
\label{eq:localHQ}
\end{align}
and therefore, the DoC of the $i^\text{th}$ term is now  given by the number of non-zero entries in the $j^\text{th}$ row of $\Omega^{(i)}$, defined as $\varphi^{(i)}_j$. 
This implies that the number of gates required to implement Eq.~\eqref{eq:localHQ} for all $N$ sites scales as 
\begin{align}
\CO\left(\sum_{i=1}^{N_s}\sum_{j = 1}^{d_{(i)}} 2^{n_q \varphi^{(i)}_j}\right) \, ,
\end{align}
which is exponential in $\varphi_{j}^{(i)}$. 

To recap, in the original operator basis, the DoC of $H_\CQ$ is given by
\begin{align}
    \rm DoC &= \max \left[\rm DoC\left[f; 1\right], \rm DoC \left[F; N\right] \right] \nonumber \\
    &= \min \left[N, \deg(F) \right],
\end{align}
while in the rotated basis it is
\begin{align}
    \text{DoC} &= \max \left[\text{DoC} \left[f; \max(\varphi_j^{(i)})\right], \text{DoC} \left[F; N_S\right] \right] ,
\end{align}
where $\varphi_{j}^{(i)}$ is a measure of the sparcity of $\Omega^{(i)}$. Considering the case where both $f$ and $F$ are non-polynomial functions and thus have $\deg(f)=\deg(F)=\infty$, the DoC in the rotated basis is
\begin{align}
    \text{DoC} &= \max \left[ \max(\varphi_j^{(i)}), N_S \right].
\end{align}
This implies that the exponential volume scaling can be reduced to polynomial if one finds a basis such that $\max \left[ \max(\varphi_j^{(i)}), N_S \right] = \CO(\log_2 N)^p$ for some power $p$; the same is true for $H_\CP$ as it has the same structure as $H_\CQ$. The following section will prove that it is always possible to achieve this tuning.

\subsection{Weaved rotation matrices}
\label{ssec:construction_weaved}

This section will discuss how to construct rotation matrices which satisfy the criteria
\begin{align}
\varphi_j\leq \lceil \log_2 d\rceil+1\,.
\end{align}
Throughout this section it is generally assumed that each sub-block of $\CW$ is of similar size, $d_{(i)} \sim d$, and also that every row of each sub-block has a comparable number of non-zero elements, $\varphi^{(i)}_j \sim \varphi$. These assumptions, combined with the above criteria, imply that the number of gates needed to implement Eq.~\eqref{eq:HQRot} is
\begin{align}
\CO\left(N^{n_q} + N \left(\frac{N}{\log_2 N}\right)^{n_q}\right) \, ,
\label{eq:gate_scaling}
\end{align}
which scales polynomially, not exponentially, with the system size. 
As $H_\CP$ has the same structure, it will have the same reduction in the number of gates needed to implement its complex exponential. 
Note that the exact choice of $N_S$, \ie whether $\log_2 N$ is rounded up or down, will have $\CO(1)$ effects on the number of gates; this can be readily optimized.

Let the matrix $\Omega^{(i)}_{M}$ be any orthogonal matrix of dimension $M$ whose first column has all entries equal to $1/\sqrt{M}$; the index $(i)$ is used to denote a specific matrix as $\Omega_M$ of size $M$ is not unique. 
Some examples of such matrices are 
\begin{align}
\Omega^{(a)}_4 &=\begin{pmatrix}
  \frac{1}{2} & -\frac{1}{\sqrt{2}} & -\frac{1}{2} & 0 \\
 \frac{1}{2} & \frac{1}{\sqrt{2}} & -\frac{1}{2} & 0 \\
 \frac{1}{2} & 0 & \frac{1}{2} & -\frac{1}{\sqrt{2}} \\
 \frac{1}{2} & 0 & \frac{1}{2} & \frac{1}{\sqrt{2}}
 \end{pmatrix}\nonumber
 \\
 \Omega^{(b)}_4 &= \frac{1}{2}\begin{pmatrix}
 1 & 1 & 1 & 1 \\
 1 & 1 & -1 & -1 \\
 1 & -1 & -1 & 1 \\
 1 & -1 & 1 & -1 
 \end{pmatrix}\,.
\end{align}
This section proves that there is a subset of the matrices $\Omega_M$ that obey 
\begin{align}
\label{eq:WeavingRequire}
\eta(\Omega_M)=\lceil \log_2(M) \rceil + 1 \quad \forall \quad  M \in \mathbb{Z}^+
\,.
\end{align}
Here $\eta(\mathbb{M})$ is the sparsity of the matrix $\mathbb{M}$, defined as the largest number of non-zero entries that appear in any individual row of the matrix $\mathbb{M}$, \textit{i.e.} $\eta(\mathbb{M})=\text{max}_{j}(\varphi_j(\mathbb{M}))$, where $\varphi_j(\mathbb{M})$ was previously defined as the number of non-zero entries in the $j^\text{th}$ row of a matrix $\mathbb{M}$. 
For example, in the case above, 
\begin{align}
\eta(\Omega_4^{(a)}) = 3 \qquad \eta(\Omega_4^{(b)}) = 4
\,.
\end{align}
With this definition, $\Omega^{(a)}_4$ obeys Eq.~\eqref{eq:WeavingRequire} while $\Omega^{(b)}_4$ does not. 
Matrices that obey Eq.~\eqref{eq:WeavingRequire} will be called weaved matrices, $W_M$, where the subscript again labels the dimension of the square orthogonal matrix. 
The reason behind this naming convention will be made clear when their explicit construction is discussed.
The proof will proceed in two parts. 
The first focuses on constructing weaved matrices of dimensions $M = 2^m, m \in \mathbb{Z}^{0+}$, while the second focuses on constructing matrices of all other dimensions. 

For the first part, let $T^{(i,j)}_{M}(\theta)$ be a rotation matrix that performs a 2-dimensional rotation by an angle $\theta$ in an $M$-dimensional vector space in the plane spanned by the $i^\text{th}$ and $j^\text{th}$ component of the vector space.
In other words, one can write
\begin{align}
\label{eq:Tmatrix}
    T^{(i,j)}_M(\theta) &=\begin{pmatrix}
     \mathbbm{1}_{i-1} & 0 & 0 & 0 & 0 \\
     0 & \cos\theta & 0 & -\sin\theta & 0  \\
     0  & 0 & \mathbbm{1}_{j-i-1} & 0 & 0  \\
     0 & \sin\theta & 0 & \cos\theta & 0 \\
     0 & 0 & 0 & 0 & \mathbbm{1}_{M-j}\\
\end{pmatrix}
\,,
\end{align}
where $\mathbbm{1}_i$ is the $i\times i$ identity matrix.
A few relevant examples are
\begin{align}
T^{(1,2)}_2(\theta) &=\begin{pmatrix}\cos \theta& -\sin \theta\\
\sin \theta & \cos \theta 
\end{pmatrix}\,, \nonumber\\
T^{(1,3)}_4(\theta) &=\begin{pmatrix}\cos \theta&0& -\sin \theta&0\\
0&1&0&0 \\
\sin \theta & 0& \cos \theta & 0\\
0&0&0&1
\end{pmatrix} \,.
\end{align}

This matrix is used to create $W_{2^m}$, assuming that $W_{2^{m-1}}$ has already been constructed. In particular, $T$ weaves together two copies of $W_{2^{m-1}}$ via
\begin{align}
W_{2^m} = \begin{pmatrix}W_{2^{m-1}}& 0\\
0 & W_{2^{m-1}} 
\end{pmatrix}\times T^{(1, 1+2^{m -1})}_{2^m}\left(\pi/4\right)\,,
\label{eq:Weaved2n}
\end{align}
Notice that due to the form of the matrix $T$, explicitly shown in Eq.~\eqref{eq:Tmatrix}, its effect is to introduce a single non-zero entry into each row of the off-diagonal sub-blocks.
For example, 
\begin{align}
W_4 = \begin{pmatrix}
\frac{W_2^{(11)}}{\sqrt{2}} & W_2^{(12)} & -\frac{W_2^{(11)}}{\sqrt{2}} & 0 \\
\frac{W_2^{(21)}}{\sqrt{2}} & W_2^{(22)} & -\frac{W_2^{(21)}}{\sqrt{2}} & 0 \\
\frac{W_2^{(11)}}{\sqrt{2}} & 0 & \frac{W_2^{(11)}}{\sqrt{2}} & W_2^{(12)} \\
\frac{W_2^{(21)}}{\sqrt{2}} & 0 & \frac{W_2^{(21)}}{\sqrt{2}} & W_2^{(22)} 
\end{pmatrix}
\end{align}
Therefore, the $\eta$ parameter of the matrix $W_{2^m}$ is given by
\begin{align}
\eta(W_{2^m}) = \eta(W_{2^{m-1}}) +1
\,.
\end{align}
However, $W_{2^{m-1}}$ can again be created by weaving together two copies of $W_{2^{m-2}}$ using an analogous $T$ matrix, with $m \rightarrow m-1$. 
By iteration one therefore finds
\begin{align}
\eta(W_{2^{m}}) &= \eta(W_{2^0}) + m \, .
\end{align}
Since $W_{2^0} = 1$, it follows that
\begin{align}
\eta(W_{2^{m}}) = 1+ m
\,.
\end{align}
Therefore $W_{2^m}$ satisfies Eq.~\eqref{eq:WeavingRequire} for any non-zero integer $m$. The orthogonality of $W_{2^m}$ can also be proven inductively, as $W_{2^m}$ is orthogonal as long as $W_{2^{m-1}}$ is due to the matrix $T$ being a rotation matrix.
The first few weaved matrices are given explicitly by
\begin{align}
W_1 &= 1, \nonumber\\
W_2 &= \frac{1}{\sqrt{2}}\begin{pmatrix}1& -1 \\
1 & 1 \end{pmatrix}, \nonumber\\
W_4&=\begin{pmatrix}
  \frac{1}{2} & -\frac{1}{\sqrt{2}} & -\frac{1}{2} & 0 \\
 \frac{1}{2} & \frac{1}{\sqrt{2}} & -\frac{1}{2} & 0 \\
 \frac{1}{2} & 0 & \frac{1}{2} & -\frac{1}{\sqrt{2}} \\
 \frac{1}{2} & 0 & \frac{1}{2} & \frac{1}{\sqrt{2}}
 \end{pmatrix}.
\end{align}
To summarize, the results presented so far show that it is possible to construct an orthogonal matrix of dimension $2^m$, with $m$ any positive integer, whose first column has all entries equal to $1/\sqrt{2^m}$ and
\begin{align}
\eta(W_{2^m}) = m+1 \, .
\end{align}

The second part of the proof shows how to construct weaved matrices for general values of $M \neq 2^m$. 
This makes use of the fact that any integer can be written in a binary form with $K = \lceil \log_2(M)\rceil$ bits, such that one can define a set $\tilde{m}(M)$ containing the position of the non-zero bits. 
For example, $\tilde m(5) = \{0,2\}$, while $\tilde m(10) = \{1, 3\}$. Furthermore, $\tilde{m}_i$ denotes the $i^{\rm {th}}$ entry in $\tilde{m}(M)$.
With this, $W_M$, for $M \neq 2^m$, is defined to be
\begin{align}
W_{M} &= \begin{pmatrix}
W_{2^{\tilde{m}_{1}}}&0&0\\
0&\ddots&0 \\
0&0& W_{2^{\tilde{m}_{k}}}
\end{pmatrix}  \times \prod_{j = 1}^{k-1}T^{\left(1, 1+b_j(\tilde m)\right)}_M(\Theta_j) \,,
\label{eq:WeavedRotGen}
\end{align}
where $k \equiv \dim \tilde m(M)$ and
\begin{align}
    b_j(\tilde m) = \sum_{i = 1}^{j}2^{\tilde{m}_i}
    \,,
\end{align}
which is the binary representation that only keeps the first $j$ digits. 
The matrix multiplication of the product reads left to right, $T^{(i,j)}_M$ is the same rotation matrix defined above and $\theta_j$ is given by
\begin{align}
\theta_j = \arccos \left[\left(1+\left[2^{\tilde{m}_{j+1}}/b_j(\tilde m)\right]\right)^{-1/2}\right]
\,.
\end{align}
Note that the product of $T$ matrices can be simplified to 
\begin{align}
& \prod_{j = 1}^{k-1}T^{\left(1, 1+b_j(\tilde m)\right)}_M(\Theta_j) = \\ &\begin{psmallmatrix}\scriptstyle\prod_{j = 1}^{k-1}\cos(\Theta_j)&0&-\sin\Theta_1&0&\cdots&-\sin\Theta_{k-1}&0\\
 0&\mathbbm{1}_{\Delta_{1}}&0&0&\cdots &0&0\\
 \sin \Theta_1 &0&\cos \Theta_1 &0 & \cdots& 0 & 0\\
 0 &0 &0 & \mathbbm{1}_{\Delta_{2}} &\cdots&0&0\\
 \vdots &\vdots & \vdots&\vdots &\ddots&0&0\\
 \sin \Theta_{k-1}&0&0&0&\cdots & \cos \Theta_k & 0 \\
 0&0&0&0&\cdots &0&  \mathbbm{1}_{2^{\tilde m_k}-1} \nonumber
  \end{psmallmatrix}\,, 
  \label{eq:TProdSimp}
\end{align}
where $\Delta_{i}$ measures the difference between the dimension of `neighboring' $W_{2^{\tilde m_i}}$,
\begin{align}
\Delta_i \equiv 2^{\tilde m_{i+1}}  -2^{\tilde m_i}-1 \, .
\end{align}
Note that because $W_M$ is constructed from products of orthogonal matrices, $W_M$ is itself orthogonal. 

It will now be shown that $W_M$ obeys the sparsity constraint
\begin{align}
\eta(W_M) = \lceil \log_2(M) \rceil + 1 \, .
\end{align}
As shown in the first part of the proof, the sparsity of each of the $W_{2^{\tilde{m}_i}}$ sub-blocks is given by $1+\tilde m_i$. 
Each of these blocks is multiplied non-trivially by at most $\lfloor \log_2 M \rfloor - \tilde m_i+1$ $T$-matrices from the right, each of which increases the sparsity by one. 
Furthermore, the largest $\tilde m_i$ is given by 
\begin{align}
    \tilde m_k = \lfloor \log_2 M \rfloor
    \,,
\end{align}
and its corresponding $W_{2^{\tilde m_k}}$ is multiplied non-trivially by exactly one $T$-matrix. Therefore
\begin{align}
\eta(W_M) &= 1 + \left(\tilde m_k +1\right)
\nonumber\\
&= 1 + \lceil \log_2 M \rceil
\,,
\end{align}
for any value of $M$.

\begin{figure*}[t]
\subfloat[\label{fig:OriginalBasis} Degree of coupling in the original basis. The system has maximum DoC, resulting in circuit depths that are exponential in the volume.]{\includegraphics[width=0.48
\textwidth]{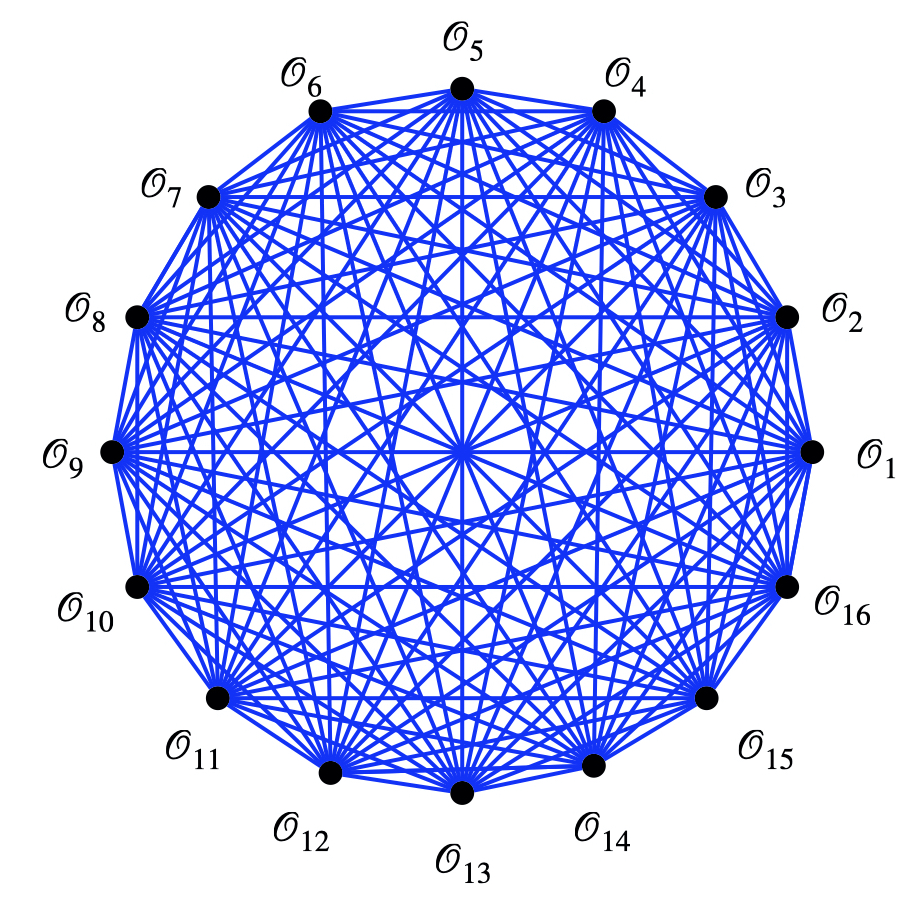}}
\hfill
\subfloat[\label{fig:WeavedBasis}Degree of connectivity in the weaved basis. This reduction in connectivity results in a circuit depth that scales polynomially in the volume.]{\includegraphics[width=0.42\textwidth]{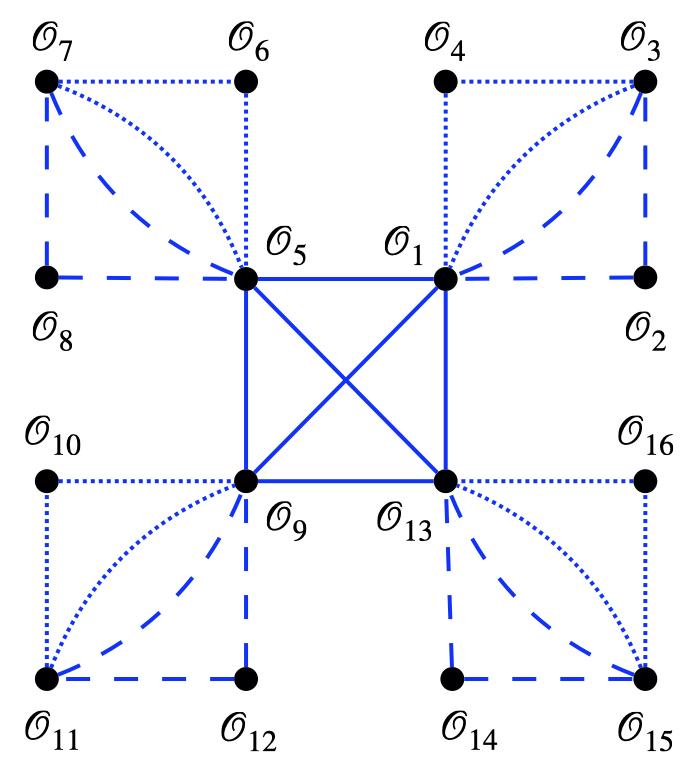}}
\caption{Diagrammatic depiction of the Degree of Connectivity for a system with 16 total operators. Fig.~\ref{fig:OriginalBasis} shows the result for a Hamiltonian with the original constraint giving rise to maximal  ${\rm DoC} = 16$. Fig.~\ref{fig:WeavedBasis} illustrates the DoC after the operator basis has been transformed to reduce the degree of connectivity, using the weaved operation. The degree of connectivity is now $\rm DoC = 4$, coming from the terms involving $\CO_1$, $\CO_5$, $\CO_9$ and $\CO_{13}$. Note that both Hamiltonians have the same spectrum, up to digitization effects. The operator $\CO_i$ could be $\CP_i$ or $\CQ_i$.}
\label{fig:BasisRed}
\end{figure*}

To summarize, it is always possible to construct an orthogonal matrix of dimension $M$ whose first column has all entries equal to $1/\sqrt{M}$ and whose rows contain at most $\lceil \log_2 M \rceil + 1$ non-zero entries.  It is instructive to present a visualization of the dramatic change in the DoC if the procedure described in the previous section is carried out using weaved matrices. 
In Fig.~\ref{fig:BasisRed}, the DoC for each operator in a system that has a total of sixteen operators is presented. 
Specifically,  Fig.~\ref{fig:OriginalBasis} shows the coupling between operators in the original basis. 
Note that this is a maximally coupled system. On the other hand, Fig.~\ref{fig:WeavedBasis} shows the coupling between sites in the weaved basis, which has a significant reduction. 

\subsection{Classical Computational Cost of Constructing Weaved Matrices}
\label{sec:ClassicalCalc}
In this section, a derivation of the classical computation cost required to construct weaved matrices is given.
It will show that the classical cost required to construct the weaved matrix $W_M$ is $\CO(M \log_2 M)$.

Recall that the naive cost of multiplying together two arbitrary square matrices of dimension  $M$ is $\CO(M^3)$. 
This can be seen by noting that each of the $\CO(M^2)$ entries in the new matrix requires $\CO(M)$ calculations. 
Note that there are advanced algorithms that reduce the scaling to less than cubic \cite{Strassen1969, Alman2010}; however, this section will assume that the computational cost scales like the total number of simple algebraic manipulations. 

The computational cost of multiplying matrices is reduced if the matrices are sparse, as the zero entries can simply be ignored.
Therefore, if the matrices are exceptionally sparse, the cost can be dramatically reduced.
The cost of constructing a weaved matrix of dimension $W_{2^m}$ with $m$ a positive integer greater than 1 is derived first.
By construction this matrix  has only $2^m(m+1)$ non-zero entries and furthermore, the first column has all $2^m$ entries non-zero and therefore the remainder of the matrix contains only $2^{m}m$ non-zero entries. 
To estimate the computational cost of constructing this matrix, recall that the construction of $W_{2^m}$ is done by weaving together two copies of $W_{2^{m-1}}$, as given in Eq.~\eqref{eq:Weaved2n}. 
This means that the $2^{m-1}(m-1)$ non-zero entries in each copy of $W_{2^{m-1}}$, excluding the first column, are multiplied by the identity matrix, requiring $2^{m}(m-1)$ algebraic multiplications. 
Next, each first column of $W_{2^{m-1}}$ is multiplied by $\cos \pi/4$ and $\pm \sin \pi/4$, where the sign on the $\sin \pi/4$ depends on whether it is the lower or upper $W_{2^{m-1}}$. 
This requires $2 \times 2 \times 2^{m-1}$ algebraic multiplications. Lastly, due to the structure of $T$ shown in Eq.~\eqref{eq:Tmatrix}, these two procedures factorize and so there is no addition that has to happen (unlike when multiplying generic matrices). 
Therefore, the total number of algebraic manipulations to construct $W_{2^m}$, assuming that $W_{2^{m-1}}$ is already constructed, is $2^m(m+1)$, which is equal to the total number of non-zero entries in $W_{2^m}$. 
If $W_{2^{m-1}}$ is not constructed, then it will take $2^{m-1}m$ algebraic manipulations. 
To construct $W_{2^m}$, starting from $W_1$ thus requires
\begin{align}
\sum_{j = 1}^{m}2^j(j+1) = m 2^{m+1}
\end{align}
algebraic manipulations. 
The total classical computational cost of constructing $W_{2^m}$ is therefore $\CO(m 2^m)$.

Similar arguments can be used to understand the scaling of constructing $W_{M}$, for $M$ an arbitrary positive integer. 
Constructing all of the necessary $W_{2^{\tilde m_i}}$ will take $\CO(M \log_2 M)$ as the maximum $\tilde m$ is given by $\lfloor \log_2 M \rfloor$. 
The construction of $W_{M}$ is again constructed by multiplying with the product of $T$ matrices shown in Eq.~\eqref{eq:TProdSimp}. 
Repeating similar steps as before still results in the total computational cost of computing $W_M$ scaling as $\CO(M \log_2 M)$.

\subsection{Comments on generalizations of this approach}

Note that the discussion so far assumed that the two operators obeyed the canonical commutation relations.
This fact was important to ensure that the bases of $\CQ'_i$ and $\CP'_i$ are still related by a Fourier transform, and was not material in reducing the DoC between the operators. 
One therefore expects that similar `change of basis' methods would also work for Hamiltonians which are constructed out of operators that do not satisfy canonical commutation relations; this would require finding an efficient implementation of the new relation between the two basis $\CQ'_i$ and $\CP'_i$.

Besides this, there are a few important caveats about the applicability and usefulness of this method applied to these other cases. 
The first, most important caveat is that, as currently written, this method is only exact when applied to operators that have an infinite dimensional spectrum. 
This can be understood by realizing that when performing Kronecker sums of operators, the dimensionality of the space that they span is increased. 
However, if this space is already infinite, it does not make a difference. 
The generalization and application to discrete operators has interesting applications and will be carried out in future work. 

Second, for polynomial functions $F$ and $G$ of order $p$, no term contains more than $p$ operators multiplied together. 
Thus, the DoC is in fact no larger than the order of the polynomial $p$, and the number of gates required\footnote{The case with $p = 1$ and $p =2$ was shown in Ref.~\cite{PhysRevA.99.052335}; the generalization to arbitrary $p$ can be understood by counting the number of individual terms in $\bigotimes_{i=1}^p x_i$, with $x = \frac{x_\text{max}}{2^{n_q}-1} \sum_{j=1}^{n_q} 2^{j} \sigma^z_j$.} is $\CO(n_q^p)$. 
However, if $p$ is large, it still might be beneficial to utilize the change of basis in order to decrease the coefficient in front of the polynomial scaling. 

\section{Application to Compact U(1) Gauge Theory}
\label{sec:Uone}

An example of where this change of basis is of fundamental importance is for an implementation of the full gauge-fixed compact 2+1 dimensional pure U(1) gauge theory. 
In this section, after briefly reviewing the formulation, the change of basis is performed and the dramatic reduction in the number of gates needed to implement time evolution of this Hamiltonian using Suzuki-Trotter methods is discussed.

\subsection{Review of lattice formulation of compact U(1)  gauge theory}
\label{sec:Formulation}

The Hamiltonian of a  \uone gauge theory in two spatial dimensions is given by
\begin{align}
    H = \int {\rm d}^2x \left( \vec{E}(x)^2 + B(x)^2\right)
    \,,
\end{align}
where the electric and magnetic fields $\vec{E}(x)$ and $B(x)$, respectively, are related to the vector potential via
\begin{align}
    \vec{E}(x) &= \partial_t \vec{A}(x) \nonumber\\
    B(x) &= \vec{\nabla} \times \vec{A}(x)
    \,.
\end{align}
Note that in two dimensions the curl of a vector field yields a pseudoscalar, while in the more familiar three dimensions it yields a vector. 
In a theory without matter, as considered here, Gauss' law gives a constraint on the electric field $\vec\nabla \cdot \vec E = 0$.

Putting the theory onto a spatial lattice, the Hamiltonian can still be split into a contribution from electric and magnetic fields
\begin{align}
H = H_E + H_B\,,
\end{align}
where electric fields are defined on the links that connect the lattice sites, and magnetic fields are defined on plaquettes, which are constructed out of four links. 
The Hamiltonian can be written in a gauge-redundancy free manner by going into the dual basis formulation \cite{Drell:1978hr,Kaplan:2018vnj, Haase:2020kaj, Bender:2020ztu}. 
In particular, the electric links are exchanged for rotor variables, which are plaquette variables, via
\begin{align}
\vec E_\ell = \vec \nabla_L \times R_p\,,
\end{align}
where $\vec\nabla_L$ is the differential operator, defined on a lattice. The magnetic field is left unaltered, defined on the plaquettes as $B_p$.   
In a quantum theory, $R_p$ and $B_p$ are promoted to operators, which satisfy the canonical commutation relations
\begin{align}
 \left[ \hat B_p, \hat R_{p'} \right] = i \,  \delta_{pp'}
    \,.
\end{align}

In a pure gauge theory with no charges, the (local) Gauss law constraint
\begin{align}
\vec{\nabla}_L \cdot \vec{\hat{E}}_\ell = 0
\end{align}
is automatically satisfied. 
However, for periodic boundary conditions, there is a global constraint that can be thought of as a magnetic Gauss law, stating that the total magnetic field through a closed surface vanishes
\begin{align}
\sum_p \hat{B}_p = 0
\,.
\end{align}
Putting this information together, the electric Hamiltonian is given by
\begin{align}
\hat{H}_E = \frac{g^2}{2a}\sum_{p=1}^{N_p}
     (\vec{\nabla}_L \times \hat R_p) 
     \,.
     \label{eq:HR}
\end{align}
where $g$ is the dimensionless gauge coupling; note that this Hamiltonian is bilinear and also nearest neighbor in the rotor fields. 
Note that $N_p$  is the number of independent plaquettes, given by $N_p = N_x N_y-1$ where $N_{x(y)}$ is the number of lattice sites in the $\hat{x}(\hat y)$ direction. 
The magnetic Hamiltonian, for the compact version of the theory is given by
\begin{align}
 \hat{H}_B =  -\frac{1}{2a\, g^2}\left(\sum_{p=1}^{N_p}\cos  \left[\hat B_p\right]+\cos\left[\sum_{p=1}^{N_p} \hat B_p\right]\right)
    \,,
    \label{eq:HB}
\end{align}
where an overall constant shift has been dropped. 

The implementation of this Hamiltonian onto a digital quantum computer requires the introduction of a digitization and truncation scheme. 
This was done in Ref.~\cite{Bauer_2021}, but the exact details are not necessary for carrying out the change of basis procedure described in Sec.~\ref{ssec:GenCoB}. %
The only relevant detail is that in this scheme, the operators $\hat{B}_p$ and $\hat{R}_p$ have equally spaced eigenvalues and their eigenbases are related by Fourier transform. 
Additionally, since the functions $f$ and $F$ are periodic, care needs to be taken that eigenvalues of the new operators do not exceed the range $-\pi$ to $\pi$.

Before implementing the change of basis procedure, it is again important to highlight that there are many other proposed Hamiltonian formulations and computational strategies for simulating U(1) (or other gauge theories) on a quantum computer \cite{Kaplan:2018vnj, Unmuth-Yockey:2018xak, Haase:2020kaj, Bender:2020ztu, Anishetty:2009nh,  Raychowdhury:2018osk, Raychowdhury:2019iki, Anishetty:2009ai, Manu2010, Manu2011, Raychowdhury:2013rwa, Zohar2015, Zohar2017, Alexandru:2019nsa, Ji2020, Hackett2019,kreshchuk2020quantum, Kreshchuk2021,Buser:2020cvn, Kaplan:2002wv, PhysRevD.104.094519,  Clemente:2022cka, Carena:2022hpz}.
Not all of these proposals depend on a fully gauge-fixed Hamiltonian. 
Instead, they utilize other methods for ensuring that Gauss' Laws are ultimately satisfied. 
It is currently an open question of what approach is best, and in fact, the answer may change as quantum computers evolve beyond the Noisy-Intermediate Scale Quantum (NISQ)-era ~\cite{Preskill2018quantumcomputingin}. 
The purpose of this section is to demonstrate how one of the main drawbacks of using fully gauge-fixed Hamiltonians, the high DoC, may be ameliorated.

\subsection{Utilizing the Change of Basis Procedure}

The Hamiltonian for the compact \uone theory presented above is of the general form given in Eq.~\eqref{eq: ArbitraryHam} with
\begin{align}
f(x) = F(x) = \cos(x) \qquad g(x) = G(x) = x^2
\,.
\end{align}
As $g$ and $G$ are both quadratic in the operators, the number of gates needed to implement their time evolution scales quadratically with $n_q$, the number of qubits used to represent each operator. 
Because there are $N_p$ such terms, the number of gates required to implement $g$ and $G$ is $\CO(N_p n_q^2)$.
However, as both $f$ and $F$ are non-polynomial functions, the gates needed to implement their time evolution will naively scale exponentially with the number of qubits each operator acts on. 
For all $N_p$ of the $f$ terms the scaling would then be $\CO(2^{n_q})$, leading to a total scaling of $\CO(N_p 2^{n_q})$. 
The function $F$ acts on the entire system of qubits, and therefore scales as $\CO(2^{n_q N_p})$, which is exponential in $N_p$. 
However, this naive scaling can be reduced using the methods developed in Sec.~\ref{sec:CoB}, leading to a dramatic reduction of the number of gates needed. 

Before providing some numerical examples of the improvement in scaling, note that carrying out a change of basis in the magnetic Hamiltonian will alter the electric Hamiltonian. 
Because $H_E$ consists of only bilinear terms, performing a basis change can introduce up to a maximum of $N_p^2$ such terms, leading to a worst case scaling of $\CO(N_p^2 n_q^2)$ in the weaved basis. 
However, a potential growth in cost in the electric Hamiltonian will be more than offset by the breaking of the exponential scaling in the magnetic Hamiltonian.

To get a sense of the full scope of this improvement, one can estimate the number of gates necessary for implementing the original versus weaved Hamiltonians.
Using $\CO( 2^{n_q N})$ scaling, the original operator basis requires $\CO(10^5)$ gates for a $3 \times 3$ lattice and $\CO(10^9)$ gates for a $4 \times 4$ lattice, assuming two qubits per operator. 
In the weaved basis, using Eq.~\eqref{eq:gate_scaling}, both of these lattices require only $\CO(10^2)$ gates.

\section{Conclusion}
\label{sec:concl}

This paper presented a general method to reduce the degree of coupling of a Hamiltonian of the form
\begin{align}
H &= \sum_{i=1}^N\left( f \left[\CQ_i\right]+g \left[\CP_i\right]\right) + F\left[\sum_{i= 1}^N \CQ_i \right] +G\left[\sum_{i= 1}^N \CP_i \right] \,,
\end{align}
which arises in systems where the Hamiltonian is given by a sum over functions of individual operators, with a global constraint on the sum of these operators. 
A well known example of such a system is a compact \uone gauge theory that is formulated on a lattice with periodic boundary conditions. 

The method requires performing a change of operator basis, which reduces to maximal number of terms involved in the arguments of a non-polynomial function from $N$ to $\log_2 N$.
The matrices defining the operator redefinition are constructed using iterative methods by weaving together  lower dimensional matrices in a particular way.
These matrices can be constructed very efficiently, with the computational cost scaling as $N \log_2 N$. 

The presented technique breaks the computational cost required to simulate such a theory from scaling exponentially with the system size $N$ to scaling polynomial with the system size. 
The expectation is that this method will have applications beyond the example of a \uone theory which was the main motivation for this work. A detailed study of the implementation of the \uone theory is left for future work.

\begin{acknowledgments}
The authors would like to thank Hank Lamm, Natalie Klco, Jesse Stryker, Tobias Osborne and Zhiyao Li for useful discussions and comments on the manuscript. CWB and BN are supported by the U.S. Department of Energy (DOE), Office of Science under contract DE-AC02-05CH11231. In particular, support comes from Quantum Information Science Enabled Discovery (QuantISED) for High Energy Physics (KA2401032). DMG acknowledges the support of the CERN Quantum Technology Initiative as well as thanking the Mainz Institute for Theoretical Physics (MITP) of the Cluster of Excellence PRISMA+ (Project ID 39083149) for its hospitality and support during the writing of this manuscript. CK is supported by the DOE Computational Science Graduate Fellowship under award number DE-SC0020347. This research benefited from the resources of the Oak Ridge Leadership Computing Facility, which is a DOE Office of Science User Facility supported under Contract DE-AC05-00OR22725
 \end{acknowledgments}

\bibliographystyle{apsrev4-1}
\bibliography{references}

\begin{thebibliography}{53}%
\makeatletter
\providecommand \@ifxundefined [1]{%
 \@ifx{#1\undefined}
}%
\providecommand \@ifnum [1]{%
 \ifnum #1\expandafter \@firstoftwo
 \else \expandafter \@secondoftwo
 \fi
}%
\providecommand \@ifx [1]{%
 \ifx #1\expandafter \@firstoftwo
 \else \expandafter \@secondoftwo
 \fi
}%
\providecommand \natexlab [1]{#1}%
\providecommand \enquote  [1]{``#1''}%
\providecommand \bibnamefont  [1]{#1}%
\providecommand \bibfnamefont [1]{#1}%
\providecommand \citenamefont [1]{#1}%
\providecommand \href@noop [0]{\@secondoftwo}%
\providecommand \href [0]{\begingroup \@sanitize@url \@href}%
\providecommand \@href[1]{\@@startlink{#1}\@@href}%
\providecommand \@@href[1]{\endgroup#1\@@endlink}%
\providecommand \@sanitize@url [0]{\catcode `\\12\catcode `\$12\catcode
  `\&12\catcode `\#12\catcode `\^12\catcode `\_12\catcode `\%12\relax}%
\providecommand \@@startlink[1]{}%
\providecommand \@@endlink[0]{}%
\providecommand \url  [0]{\begingroup\@sanitize@url \@url }%
\providecommand \@url [1]{\endgroup\@href {#1}{\urlprefix }}%
\providecommand \urlprefix  [0]{URL }%
\providecommand \Eprint [0]{\href }%
\providecommand \doibase [0]{http://dx.doi.org/}%
\providecommand \selectlanguage [0]{\@gobble}%
\providecommand \bibinfo  [0]{\@secondoftwo}%
\providecommand \bibfield  [0]{\@secondoftwo}%
\providecommand \translation [1]{[#1]}%
\providecommand \BibitemOpen [0]{}%
\providecommand \bibitemStop [0]{}%
\providecommand \bibitemNoStop [0]{.\EOS\space}%
\providecommand \EOS [0]{\spacefactor3000\relax}%
\providecommand \BibitemShut  [1]{\csname bibitem#1\endcsname}%
\let\auto@bib@innerbib\@empty
\bibitem [{\citenamefont {Dirac}(1964)}]{Dirac1964}%
  \BibitemOpen
  \bibfield  {author} {\bibinfo {author} {\bibfnamefont {P.~A.~M.}\
  \bibnamefont {Dirac}},\ }\href@noop {} {\emph {\bibinfo {title} {Lectures on
  quantum mechanics}}}\ (\bibinfo {year} {1964})\BibitemShut {NoStop}%
\bibitem [{\citenamefont {Henneaux}\ and\ \citenamefont
  {Teitelboim}(1992)}]{Henneaux1992}%
  \BibitemOpen
  \bibfield  {author} {\bibinfo {author} {\bibfnamefont {M.}~\bibnamefont
  {Henneaux}}\ and\ \bibinfo {author} {\bibfnamefont {C.}~\bibnamefont
  {Teitelboim}},\ }\href {http://www.jstor.org/stable/j.ctv10crg0r} {\emph
  {\bibinfo {title} {Quantization of Gauge Systems}}}\ (\bibinfo  {publisher}
  {Princeton University Press},\ \bibinfo {year} {1992})\BibitemShut {NoStop}%
\bibitem [{\citenamefont {Bojowald}(2010)}]{bojowald_2010}%
  \BibitemOpen
  \bibfield  {author} {\bibinfo {author} {\bibfnamefont {M.}~\bibnamefont
  {Bojowald}},\ }\href {\doibase 10.1017/CBO9780511921759} {\emph {\bibinfo
  {title} {Canonical Gravity and Applications: Cosmology, Black Holes, and
  Quantum Gravity}}}\ (\bibinfo  {publisher} {Cambridge University Press},\
  \bibinfo {year} {2010})\BibitemShut {NoStop}%
\bibitem [{\citenamefont {Suzuki}(1976{\natexlab{a}})}]{Suzuki:1976be}%
  \BibitemOpen
  \bibfield  {author} {\bibinfo {author} {\bibfnamefont {M.}~\bibnamefont
  {Suzuki}},\ }\href {\doibase 10.1007/BF01609348} {\bibfield  {journal}
  {\bibinfo  {journal} {Commun. Math. Phys.}\ }\textbf {\bibinfo {volume}
  {51}},\ \bibinfo {pages} {183} (\bibinfo {year}
  {1976}{\natexlab{a}})}\BibitemShut {NoStop}%
\bibitem [{\citenamefont {Suzuki}(1976{\natexlab{b}})}]{10.1143/PTP.56.1454}%
  \BibitemOpen
  \bibfield  {author} {\bibinfo {author} {\bibfnamefont {M.}~\bibnamefont
  {Suzuki}},\ }\href {\doibase 10.1143/PTP.56.1454} {\bibfield  {journal}
  {\bibinfo  {journal} {Progress of Theoretical Physics}\ }\textbf {\bibinfo
  {volume} {56}},\ \bibinfo {pages} {1454} (\bibinfo {year}
  {1976}{\natexlab{b}})},\ \Eprint
  {http://arxiv.org/abs/https://academic.oup.com/ptp/article-pdf/56/5/1454/5264429/56-5-1454.pdf}
  {https://academic.oup.com/ptp/article-pdf/56/5/1454/5264429/56-5-1454.pdf}
  \BibitemShut {NoStop}%
\bibitem [{\citenamefont {Trotter}(1959)}]{10.2307/2033649}%
  \BibitemOpen
  \bibfield  {author} {\bibinfo {author} {\bibfnamefont {H.~F.}\ \bibnamefont
  {Trotter}},\ }\href {http://www.jstor.org/stable/2033649} {\bibfield
  {journal} {\bibinfo  {journal} {Proceedings of the American Mathematical
  Society}\ }\textbf {\bibinfo {volume} {10}},\ \bibinfo {pages} {545}
  (\bibinfo {year} {1959})}\BibitemShut {NoStop}%
\bibitem [{\citenamefont {Zohar}\ and\ \citenamefont
  {Reznik}(2011)}]{PhysRevLett.107.275301}%
  \BibitemOpen
  \bibfield  {author} {\bibinfo {author} {\bibfnamefont {E.}~\bibnamefont
  {Zohar}}\ and\ \bibinfo {author} {\bibfnamefont {B.}~\bibnamefont {Reznik}},\
  }\href {\doibase 10.1103/PhysRevLett.107.275301} {\bibfield  {journal}
  {\bibinfo  {journal} {Phys. Rev. Lett.}\ }\textbf {\bibinfo {volume} {107}},\
  \bibinfo {pages} {275301} (\bibinfo {year} {2011})}\BibitemShut {NoStop}%
\bibitem [{\citenamefont {Zohar}\ \emph {et~al.}(2012)\citenamefont {Zohar},
  \citenamefont {Cirac},\ and\ \citenamefont
  {Reznik}}]{PhysRevLett.109.125302}%
  \BibitemOpen
  \bibfield  {author} {\bibinfo {author} {\bibfnamefont {E.}~\bibnamefont
  {Zohar}}, \bibinfo {author} {\bibfnamefont {J.~I.}\ \bibnamefont {Cirac}}, \
  and\ \bibinfo {author} {\bibfnamefont {B.}~\bibnamefont {Reznik}},\ }\href
  {\doibase 10.1103/PhysRevLett.109.125302} {\bibfield  {journal} {\bibinfo
  {journal} {Phys. Rev. Lett.}\ }\textbf {\bibinfo {volume} {109}},\ \bibinfo
  {pages} {125302} (\bibinfo {year} {2012})}\BibitemShut {NoStop}%
\bibitem [{\citenamefont {Hauke}\ \emph
  {et~al.}(2013{\natexlab{a}})\citenamefont {Hauke}, \citenamefont {Marcos},
  \citenamefont {Dalmonte},\ and\ \citenamefont {Zoller}}]{Hauke2013}%
  \BibitemOpen
  \bibfield  {author} {\bibinfo {author} {\bibfnamefont {P.}~\bibnamefont
  {Hauke}}, \bibinfo {author} {\bibfnamefont {D.}~\bibnamefont {Marcos}},
  \bibinfo {author} {\bibfnamefont {M.}~\bibnamefont {Dalmonte}}, \ and\
  \bibinfo {author} {\bibfnamefont {P.}~\bibnamefont {Zoller}},\ }\href
  {\doibase 10.1103/physrevx.3.041018} {\bibfield  {journal} {\bibinfo
  {journal} {Physical Review X}\ }\textbf {\bibinfo {volume} {3}} (\bibinfo
  {year} {2013}{\natexlab{a}}),\ 10.1103/physrevx.3.041018}\BibitemShut
  {NoStop}%
\bibitem [{\citenamefont {Halimeh}\ and\ \citenamefont
  {Hauke}(2020)}]{Halimeh:2019svu}%
  \BibitemOpen
  \bibfield  {author} {\bibinfo {author} {\bibfnamefont {J.~C.}\ \bibnamefont
  {Halimeh}}\ and\ \bibinfo {author} {\bibfnamefont {P.}~\bibnamefont
  {Hauke}},\ }\href {\doibase 10.1103/PhysRevLett.125.030503} {\bibfield
  {journal} {\bibinfo  {journal} {Phys. Rev. Lett.}\ }\textbf {\bibinfo
  {volume} {125}},\ \bibinfo {pages} {030503} (\bibinfo {year} {2020})},\
  \Eprint {http://arxiv.org/abs/2001.00024} {arXiv:2001.00024
  [cond-mat.quant-gas]} \BibitemShut {NoStop}%
\bibitem [{\citenamefont {Lamm}\ \emph {et~al.}(2019)\citenamefont {Lamm},
  \citenamefont {Lawrence},\ and\ \citenamefont {Yamauchi}}]{Lamm:2019bik}%
  \BibitemOpen
  \bibfield  {author} {\bibinfo {author} {\bibfnamefont {H.}~\bibnamefont
  {Lamm}}, \bibinfo {author} {\bibfnamefont {S.}~\bibnamefont {Lawrence}}, \
  and\ \bibinfo {author} {\bibfnamefont {Y.}~\bibnamefont {Yamauchi}} (\bibinfo
  {collaboration} {NuQS}),\ }\href {\doibase 10.1103/PhysRevD.100.034518}
  {\bibfield  {journal} {\bibinfo  {journal} {Phys. Rev. D}\ }\textbf {\bibinfo
  {volume} {100}},\ \bibinfo {pages} {034518} (\bibinfo {year} {2019})},\
  \Eprint {http://arxiv.org/abs/1903.08807} {arXiv:1903.08807 [hep-lat]}
  \BibitemShut {NoStop}%
\bibitem [{\citenamefont {Lamm}\ \emph {et~al.}(2020)\citenamefont {Lamm},
  \citenamefont {Lawrence},\ and\ \citenamefont {Yamauchi}}]{Lamm:2020jwv}%
  \BibitemOpen
  \bibfield  {author} {\bibinfo {author} {\bibfnamefont {H.}~\bibnamefont
  {Lamm}}, \bibinfo {author} {\bibfnamefont {S.}~\bibnamefont {Lawrence}}, \
  and\ \bibinfo {author} {\bibfnamefont {Y.}~\bibnamefont {Yamauchi}} (\bibinfo
  {collaboration} {NuQS}),\ }\href@noop {} {\  (\bibinfo {year} {2020})},\
  \Eprint {http://arxiv.org/abs/2005.12688} {arXiv:2005.12688 [quant-ph]}
  \BibitemShut {NoStop}%
\bibitem [{\citenamefont {Tran}\ \emph {et~al.}(2021)\citenamefont {Tran},
  \citenamefont {Su}, \citenamefont {Carney},\ and\ \citenamefont
  {Taylor}}]{Tran:2020azk}%
  \BibitemOpen
  \bibfield  {author} {\bibinfo {author} {\bibfnamefont {M.~C.}\ \bibnamefont
  {Tran}}, \bibinfo {author} {\bibfnamefont {Y.}~\bibnamefont {Su}}, \bibinfo
  {author} {\bibfnamefont {D.}~\bibnamefont {Carney}}, \ and\ \bibinfo {author}
  {\bibfnamefont {J.~M.}\ \bibnamefont {Taylor}},\ }\href {\doibase
  10.1103/PRXQuantum.2.010323} {\bibfield  {journal} {\bibinfo  {journal} {P.
  R. X. Quantum.}\ }\textbf {\bibinfo {volume} {2}},\ \bibinfo {pages} {010323}
  (\bibinfo {year} {2021})},\ \Eprint {http://arxiv.org/abs/2006.16248}
  {arXiv:2006.16248 [quant-ph]} \BibitemShut {NoStop}%
\bibitem [{\citenamefont {Banerjee}\ \emph {et~al.}(2012)\citenamefont
  {Banerjee}, \citenamefont {Dalmonte}, \citenamefont {Muller}, \citenamefont
  {Rico}, \citenamefont {Stebler}, \citenamefont {Wiese},\ and\ \citenamefont
  {Zoller}}]{Banerjee:2012pg}%
  \BibitemOpen
  \bibfield  {author} {\bibinfo {author} {\bibfnamefont {D.}~\bibnamefont
  {Banerjee}}, \bibinfo {author} {\bibfnamefont {M.}~\bibnamefont {Dalmonte}},
  \bibinfo {author} {\bibfnamefont {M.}~\bibnamefont {Muller}}, \bibinfo
  {author} {\bibfnamefont {E.}~\bibnamefont {Rico}}, \bibinfo {author}
  {\bibfnamefont {P.}~\bibnamefont {Stebler}}, \bibinfo {author} {\bibfnamefont
  {U.~J.}\ \bibnamefont {Wiese}}, \ and\ \bibinfo {author} {\bibfnamefont
  {P.}~\bibnamefont {Zoller}},\ }\href {\doibase
  10.1103/PhysRevLett.109.175302} {\bibfield  {journal} {\bibinfo  {journal}
  {Phys. Rev. Lett.}\ }\textbf {\bibinfo {volume} {109}},\ \bibinfo {pages}
  {175302} (\bibinfo {year} {2012})},\ \Eprint {http://arxiv.org/abs/1205.6366}
  {arXiv:1205.6366 [cond-mat.quant-gas]} \BibitemShut {NoStop}%
\bibitem [{\citenamefont {Kasper}\ \emph {et~al.}(2020)\citenamefont {Kasper},
  \citenamefont {Zache}, \citenamefont {Jendrzejewski}, \citenamefont
  {Lewenstein},\ and\ \citenamefont {Zohar}}]{Kasper:2020owz}%
  \BibitemOpen
  \bibfield  {author} {\bibinfo {author} {\bibfnamefont {V.}~\bibnamefont
  {Kasper}}, \bibinfo {author} {\bibfnamefont {T.~V.}\ \bibnamefont {Zache}},
  \bibinfo {author} {\bibfnamefont {F.}~\bibnamefont {Jendrzejewski}}, \bibinfo
  {author} {\bibfnamefont {M.}~\bibnamefont {Lewenstein}}, \ and\ \bibinfo
  {author} {\bibfnamefont {E.}~\bibnamefont {Zohar}},\ }\href@noop {} {\
  (\bibinfo {year} {2020})},\ \Eprint {http://arxiv.org/abs/2012.08620}
  {arXiv:2012.08620 [quant-ph]} \BibitemShut {NoStop}%
\bibitem [{\citenamefont {Hauke}\ \emph
  {et~al.}(2013{\natexlab{b}})\citenamefont {Hauke}, \citenamefont {Marcos},
  \citenamefont {Dalmonte},\ and\ \citenamefont {Zoller}}]{PhysRevX.3.041018}%
  \BibitemOpen
  \bibfield  {author} {\bibinfo {author} {\bibfnamefont {P.}~\bibnamefont
  {Hauke}}, \bibinfo {author} {\bibfnamefont {D.}~\bibnamefont {Marcos}},
  \bibinfo {author} {\bibfnamefont {M.}~\bibnamefont {Dalmonte}}, \ and\
  \bibinfo {author} {\bibfnamefont {P.}~\bibnamefont {Zoller}},\ }\href
  {\doibase 10.1103/PhysRevX.3.041018} {\bibfield  {journal} {\bibinfo
  {journal} {Phys. Rev. X}\ }\textbf {\bibinfo {volume} {3}},\ \bibinfo {pages}
  {041018} (\bibinfo {year} {2013}{\natexlab{b}})}\BibitemShut {NoStop}%
\bibitem [{\citenamefont {Kühn}\ \emph {et~al.}(2014)\citenamefont {Kühn},
  \citenamefont {Cirac},\ and\ \citenamefont {Bañuls}}]{Kuhn2014}%
  \BibitemOpen
  \bibfield  {author} {\bibinfo {author} {\bibfnamefont {S.}~\bibnamefont
  {Kühn}}, \bibinfo {author} {\bibfnamefont {J.~I.}\ \bibnamefont {Cirac}}, \
  and\ \bibinfo {author} {\bibfnamefont {M.-C.}\ \bibnamefont {Bañuls}},\
  }\href {\doibase 10.1103/physreva.90.042305} {\bibfield  {journal} {\bibinfo
  {journal} {Physical Review A}\ }\textbf {\bibinfo {volume} {90}} (\bibinfo
  {year} {2014}),\ 10.1103/physreva.90.042305}\BibitemShut {NoStop}%
\bibitem [{\citenamefont {Stannigel}\ \emph {et~al.}(2014)\citenamefont
  {Stannigel}, \citenamefont {Hauke}, \citenamefont {Marcos}, \citenamefont
  {Hafezi}, \citenamefont {Diehl}, \citenamefont {Dalmonte},\ and\
  \citenamefont {Zoller}}]{Stannigel:2013zka}%
  \BibitemOpen
  \bibfield  {author} {\bibinfo {author} {\bibfnamefont {K.}~\bibnamefont
  {Stannigel}}, \bibinfo {author} {\bibfnamefont {P.}~\bibnamefont {Hauke}},
  \bibinfo {author} {\bibfnamefont {D.}~\bibnamefont {Marcos}}, \bibinfo
  {author} {\bibfnamefont {M.}~\bibnamefont {Hafezi}}, \bibinfo {author}
  {\bibfnamefont {S.}~\bibnamefont {Diehl}}, \bibinfo {author} {\bibfnamefont
  {M.}~\bibnamefont {Dalmonte}}, \ and\ \bibinfo {author} {\bibfnamefont
  {P.}~\bibnamefont {Zoller}},\ }\href {\doibase
  10.1103/PhysRevLett.112.120406} {\bibfield  {journal} {\bibinfo  {journal}
  {Phys. Rev. Lett.}\ }\textbf {\bibinfo {volume} {112}},\ \bibinfo {pages}
  {120406} (\bibinfo {year} {2014})},\ \Eprint {http://arxiv.org/abs/1308.0528}
  {arXiv:1308.0528 [quant-ph]} \BibitemShut {NoStop}%
\bibitem [{\citenamefont {Drell}\ \emph {et~al.}(1979)\citenamefont {Drell},
  \citenamefont {Quinn}, \citenamefont {Svetitsky},\ and\ \citenamefont
  {Weinstein}}]{Drell:1978hr}%
  \BibitemOpen
  \bibfield  {author} {\bibinfo {author} {\bibfnamefont {S.~D.}\ \bibnamefont
  {Drell}}, \bibinfo {author} {\bibfnamefont {H.~R.}\ \bibnamefont {Quinn}},
  \bibinfo {author} {\bibfnamefont {B.}~\bibnamefont {Svetitsky}}, \ and\
  \bibinfo {author} {\bibfnamefont {M.}~\bibnamefont {Weinstein}},\ }\href
  {\doibase 10.1103/PhysRevD.19.619} {\bibfield  {journal} {\bibinfo  {journal}
  {Phys. Rev. D}\ }\textbf {\bibinfo {volume} {19}},\ \bibinfo {pages} {619}
  (\bibinfo {year} {1979})}\BibitemShut {NoStop}%
\bibitem [{\citenamefont {Kaplan}\ and\ \citenamefont
  {Stryker}(2020)}]{Kaplan:2018vnj}%
  \BibitemOpen
  \bibfield  {author} {\bibinfo {author} {\bibfnamefont {D.~B.}\ \bibnamefont
  {Kaplan}}\ and\ \bibinfo {author} {\bibfnamefont {J.~R.}\ \bibnamefont
  {Stryker}},\ }\href {\doibase 10.1103/PhysRevD.102.094515} {\bibfield
  {journal} {\bibinfo  {journal} {Phys. Rev. D}\ }\textbf {\bibinfo {volume}
  {102}},\ \bibinfo {pages} {094515} (\bibinfo {year} {2020})},\ \Eprint
  {http://arxiv.org/abs/1806.08797} {arXiv:1806.08797 [hep-lat]} \BibitemShut
  {NoStop}%
\bibitem [{\citenamefont {Haase}\ \emph {et~al.}(2021)\citenamefont {Haase},
  \citenamefont {Dellantonio}, \citenamefont {Celi}, \citenamefont {Paulson},
  \citenamefont {Kan}, \citenamefont {Jansen},\ and\ \citenamefont
  {Muschik}}]{Haase:2020kaj}%
  \BibitemOpen
  \bibfield  {author} {\bibinfo {author} {\bibfnamefont {J.~F.}\ \bibnamefont
  {Haase}}, \bibinfo {author} {\bibfnamefont {L.}~\bibnamefont {Dellantonio}},
  \bibinfo {author} {\bibfnamefont {A.}~\bibnamefont {Celi}}, \bibinfo {author}
  {\bibfnamefont {D.}~\bibnamefont {Paulson}}, \bibinfo {author} {\bibfnamefont
  {A.}~\bibnamefont {Kan}}, \bibinfo {author} {\bibfnamefont {K.}~\bibnamefont
  {Jansen}}, \ and\ \bibinfo {author} {\bibfnamefont {C.~A.}\ \bibnamefont
  {Muschik}},\ }\href {\doibase 10.22331/q-2021-02-04-393} {\bibfield
  {journal} {\bibinfo  {journal} {Quantum}\ }\textbf {\bibinfo {volume} {5}},\
  \bibinfo {pages} {393} (\bibinfo {year} {2021})},\ \Eprint
  {http://arxiv.org/abs/2006.14160} {arXiv:2006.14160 [quant-ph]} \BibitemShut
  {NoStop}%
\bibitem [{\citenamefont {Bender}\ and\ \citenamefont
  {Zohar}(2020)}]{Bender:2020ztu}%
  \BibitemOpen
  \bibfield  {author} {\bibinfo {author} {\bibfnamefont {J.}~\bibnamefont
  {Bender}}\ and\ \bibinfo {author} {\bibfnamefont {E.}~\bibnamefont {Zohar}},\
  }\href {\doibase 10.1103/PhysRevD.102.114517} {\bibfield  {journal} {\bibinfo
   {journal} {Phys. Rev. D}\ }\textbf {\bibinfo {volume} {102}},\ \bibinfo
  {pages} {114517} (\bibinfo {year} {2020})},\ \Eprint
  {http://arxiv.org/abs/2008.01349} {arXiv:2008.01349 [quant-ph]} \BibitemShut
  {NoStop}%
\bibitem [{\citenamefont {Wilson}(1974)}]{PhysRevD.10.2445}%
  \BibitemOpen
  \bibfield  {author} {\bibinfo {author} {\bibfnamefont {K.~G.}\ \bibnamefont
  {Wilson}},\ }\href {\doibase 10.1103/PhysRevD.10.2445} {\bibfield  {journal}
  {\bibinfo  {journal} {Phys. Rev. D}\ }\textbf {\bibinfo {volume} {10}},\
  \bibinfo {pages} {2445} (\bibinfo {year} {1974})}\BibitemShut {NoStop}%
\bibitem [{\citenamefont {Kogut}\ and\ \citenamefont
  {Susskind}(1975)}]{PhysRevD.11.395}%
  \BibitemOpen
  \bibfield  {author} {\bibinfo {author} {\bibfnamefont {J.}~\bibnamefont
  {Kogut}}\ and\ \bibinfo {author} {\bibfnamefont {L.}~\bibnamefont
  {Susskind}},\ }\href {\doibase 10.1103/PhysRevD.11.395} {\bibfield  {journal}
  {\bibinfo  {journal} {Phys. Rev. D}\ }\textbf {\bibinfo {volume} {11}},\
  \bibinfo {pages} {395} (\bibinfo {year} {1975})}\BibitemShut {NoStop}%
\bibitem [{\citenamefont {Banks}\ \emph {et~al.}(1976)\citenamefont {Banks},
  \citenamefont {Susskind},\ and\ \citenamefont {Kogut}}]{PhysRevD.13.1043}%
  \BibitemOpen
  \bibfield  {author} {\bibinfo {author} {\bibfnamefont {T.}~\bibnamefont
  {Banks}}, \bibinfo {author} {\bibfnamefont {L.}~\bibnamefont {Susskind}}, \
  and\ \bibinfo {author} {\bibfnamefont {J.}~\bibnamefont {Kogut}},\ }\href
  {\doibase 10.1103/PhysRevD.13.1043} {\bibfield  {journal} {\bibinfo
  {journal} {Phys. Rev. D}\ }\textbf {\bibinfo {volume} {13}},\ \bibinfo
  {pages} {1043} (\bibinfo {year} {1976})}\BibitemShut {NoStop}%
\bibitem [{\citenamefont {Polyakov}(1975)}]{Polyakov:1975rs}%
  \BibitemOpen
  \bibfield  {author} {\bibinfo {author} {\bibfnamefont {A.~M.}\ \bibnamefont
  {Polyakov}},\ }\href {\doibase 10.1016/0370-2693(75)90162-8} {\bibfield
  {journal} {\bibinfo  {journal} {Phys. Lett. B}\ }\textbf {\bibinfo {volume}
  {59}},\ \bibinfo {pages} {82} (\bibinfo {year} {1975})}\BibitemShut {NoStop}%
\bibitem [{\citenamefont {Polyakov}(1977)}]{Polyakov:1976fu}%
  \BibitemOpen
  \bibfield  {author} {\bibinfo {author} {\bibfnamefont {A.~M.}\ \bibnamefont
  {Polyakov}},\ }\href {\doibase 10.1016/0550-3213(77)90086-4} {\bibfield
  {journal} {\bibinfo  {journal} {Nucl. Phys. B}\ }\textbf {\bibinfo {volume}
  {120}},\ \bibinfo {pages} {429} (\bibinfo {year} {1977})}\BibitemShut
  {NoStop}%
\bibitem [{\citenamefont {Bullock}\ and\ \citenamefont
  {Markov}(2003)}]{Bullock2003}%
  \BibitemOpen
  \bibfield  {author} {\bibinfo {author} {\bibfnamefont {S.~S.}\ \bibnamefont
  {Bullock}}\ and\ \bibinfo {author} {\bibfnamefont {I.~L.}\ \bibnamefont
  {Markov}},\ }\href {\doibase 10.48550/ARXIV.QUANT-PH/0303039} {\  (\bibinfo
  {year} {2003}),\ 10.48550/ARXIV.QUANT-PH/0303039}\BibitemShut {NoStop}%
\bibitem [{\citenamefont {Strassen}(1969)}]{Strassen1969}%
  \BibitemOpen
  \bibfield  {author} {\bibinfo {author} {\bibfnamefont {V.}~\bibnamefont
  {Strassen}},\ }\href {\doibase 10.1007/BF02165411} {\bibfield  {journal}
  {\bibinfo  {journal} {Numerische Mathematik}\ }\textbf {\bibinfo {volume}
  {13}},\ \bibinfo {pages} {354} (\bibinfo {year} {1969})}\BibitemShut
  {NoStop}%
\bibitem [{\citenamefont {Alman}\ and\ \citenamefont
  {Williams}(2020)}]{Alman2010}%
  \BibitemOpen
  \bibfield  {author} {\bibinfo {author} {\bibfnamefont {J.}~\bibnamefont
  {Alman}}\ and\ \bibinfo {author} {\bibfnamefont {V.~V.}\ \bibnamefont
  {Williams}},\ }\href {\doibase 10.48550/ARXIV.2010.05846} {\enquote {\bibinfo
  {title} {A refined laser method and faster matrix multiplication},}\ }
  (\bibinfo {year} {2020})\BibitemShut {NoStop}%
\bibitem [{\citenamefont {Klco}\ and\ \citenamefont
  {Savage}(2019)}]{PhysRevA.99.052335}%
  \BibitemOpen
  \bibfield  {author} {\bibinfo {author} {\bibfnamefont {N.}~\bibnamefont
  {Klco}}\ and\ \bibinfo {author} {\bibfnamefont {M.~J.}\ \bibnamefont
  {Savage}},\ }\href {\doibase 10.1103/PhysRevA.99.052335} {\bibfield
  {journal} {\bibinfo  {journal} {Phys. Rev. A}\ }\textbf {\bibinfo {volume}
  {99}},\ \bibinfo {pages} {052335} (\bibinfo {year} {2019})}\BibitemShut
  {NoStop}%
\bibitem [{\citenamefont {Bauer}\ and\ \citenamefont
  {Grabowska}(2021)}]{Bauer_2021}%
  \BibitemOpen
  \bibfield  {author} {\bibinfo {author} {\bibfnamefont {C.~W.}\ \bibnamefont
  {Bauer}}\ and\ \bibinfo {author} {\bibfnamefont {D.~M.}\ \bibnamefont
  {Grabowska}},\ }\href {\doibase 10.48550/ARXIV.2111.08015} {\  (\bibinfo
  {year} {2021}),\ 10.48550/ARXIV.2111.08015}\BibitemShut {NoStop}%
\bibitem [{\citenamefont {Unmuth-Yockey}(2019)}]{Unmuth-Yockey:2018xak}%
  \BibitemOpen
  \bibfield  {author} {\bibinfo {author} {\bibfnamefont {J.~F.}\ \bibnamefont
  {Unmuth-Yockey}},\ }\href {\doibase 10.1103/PhysRevD.99.074502} {\bibfield
  {journal} {\bibinfo  {journal} {Phys. Rev. D}\ }\textbf {\bibinfo {volume}
  {99}},\ \bibinfo {pages} {074502} (\bibinfo {year} {2019})},\ \Eprint
  {http://arxiv.org/abs/1811.05884} {arXiv:1811.05884 [hep-lat]} \BibitemShut
  {NoStop}%
\bibitem [{\citenamefont {Anishetty}\ \emph {et~al.}(2010)\citenamefont
  {Anishetty}, \citenamefont {Mathur},\ and\ \citenamefont
  {Raychowdhury}}]{Anishetty:2009nh}%
  \BibitemOpen
  \bibfield  {author} {\bibinfo {author} {\bibfnamefont {R.}~\bibnamefont
  {Anishetty}}, \bibinfo {author} {\bibfnamefont {M.}~\bibnamefont {Mathur}}, \
  and\ \bibinfo {author} {\bibfnamefont {I.}~\bibnamefont {Raychowdhury}},\
  }\href {\doibase 10.1088/1751-8113/43/3/035403} {\bibfield  {journal}
  {\bibinfo  {journal} {J. Phys. A}\ }\textbf {\bibinfo {volume} {43}},\
  \bibinfo {pages} {035403} (\bibinfo {year} {2010})},\ \Eprint
  {http://arxiv.org/abs/0909.2394} {arXiv:0909.2394 [hep-lat]} \BibitemShut
  {NoStop}%
\bibitem [{\citenamefont {Raychowdhury}\ and\ \citenamefont
  {Stryker}(2020{\natexlab{a}})}]{Raychowdhury:2018osk}%
  \BibitemOpen
  \bibfield  {author} {\bibinfo {author} {\bibfnamefont {I.}~\bibnamefont
  {Raychowdhury}}\ and\ \bibinfo {author} {\bibfnamefont {J.~R.}\ \bibnamefont
  {Stryker}},\ }\href {\doibase 10.1103/PhysRevResearch.2.033039} {\bibfield
  {journal} {\bibinfo  {journal} {Phys. Rev. Res.}\ }\textbf {\bibinfo {volume}
  {2}},\ \bibinfo {pages} {033039} (\bibinfo {year} {2020}{\natexlab{a}})},\
  \Eprint {http://arxiv.org/abs/1812.07554} {arXiv:1812.07554 [hep-lat]}
  \BibitemShut {NoStop}%
\bibitem [{\citenamefont {Raychowdhury}\ and\ \citenamefont
  {Stryker}(2020{\natexlab{b}})}]{Raychowdhury:2019iki}%
  \BibitemOpen
  \bibfield  {author} {\bibinfo {author} {\bibfnamefont {I.}~\bibnamefont
  {Raychowdhury}}\ and\ \bibinfo {author} {\bibfnamefont {J.~R.}\ \bibnamefont
  {Stryker}},\ }\href {\doibase 10.1103/PhysRevD.101.114502} {\bibfield
  {journal} {\bibinfo  {journal} {Phys. Rev. D}\ }\textbf {\bibinfo {volume}
  {101}},\ \bibinfo {pages} {114502} (\bibinfo {year} {2020}{\natexlab{b}})},\
  \Eprint {http://arxiv.org/abs/1912.06133} {arXiv:1912.06133 [hep-lat]}
  \BibitemShut {NoStop}%
\bibitem [{\citenamefont {Anishetty}\ \emph {et~al.}(2009)\citenamefont
  {Anishetty}, \citenamefont {Mathur},\ and\ \citenamefont
  {Raychowdhury}}]{Anishetty:2009ai}%
  \BibitemOpen
  \bibfield  {author} {\bibinfo {author} {\bibfnamefont {R.}~\bibnamefont
  {Anishetty}}, \bibinfo {author} {\bibfnamefont {M.}~\bibnamefont {Mathur}}, \
  and\ \bibinfo {author} {\bibfnamefont {I.}~\bibnamefont {Raychowdhury}},\
  }\href {\doibase 10.1063/1.3122666} {\bibfield  {journal} {\bibinfo
  {journal} {J. Math. Phys.}\ }\textbf {\bibinfo {volume} {50}},\ \bibinfo
  {pages} {053503} (\bibinfo {year} {2009})},\ \Eprint
  {http://arxiv.org/abs/0901.0644} {arXiv:0901.0644 [math-ph]} \BibitemShut
  {NoStop}%
\bibitem [{\citenamefont {Mathur}\ \emph {et~al.}(2010)\citenamefont {Mathur},
  \citenamefont {Raychowdhury},\ and\ \citenamefont {Anishetty}}]{Manu2010}%
  \BibitemOpen
  \bibfield  {author} {\bibinfo {author} {\bibfnamefont {M.}~\bibnamefont
  {Mathur}}, \bibinfo {author} {\bibfnamefont {I.}~\bibnamefont
  {Raychowdhury}}, \ and\ \bibinfo {author} {\bibfnamefont {R.}~\bibnamefont
  {Anishetty}},\ }\href {\doibase 10.1063/1.3464267} {\bibfield  {journal}
  {\bibinfo  {journal} {Journal of Mathematical Physics}\ }\textbf {\bibinfo
  {volume} {51}},\ \bibinfo {pages} {093504} (\bibinfo {year}
  {2010})}\BibitemShut {NoStop}%
\bibitem [{\citenamefont {Mathur}\ \emph {et~al.}(2011)\citenamefont {Mathur},
  \citenamefont {Raychowdhury},\ and\ \citenamefont {Sreeraj}}]{Manu2011}%
  \BibitemOpen
  \bibfield  {author} {\bibinfo {author} {\bibfnamefont {M.}~\bibnamefont
  {Mathur}}, \bibinfo {author} {\bibfnamefont {I.}~\bibnamefont
  {Raychowdhury}}, \ and\ \bibinfo {author} {\bibfnamefont {T.~P.}\
  \bibnamefont {Sreeraj}},\ }\href {\doibase 10.1063/1.3660195} {\bibfield
  {journal} {\bibinfo  {journal} {Journal of Mathematical Physics}\ }\textbf
  {\bibinfo {volume} {52}},\ \bibinfo {pages} {113505} (\bibinfo {year}
  {2011})}\BibitemShut {NoStop}%
\bibitem [{\citenamefont {Raychowdhury}(2013)}]{Raychowdhury:2013rwa}%
  \BibitemOpen
  \bibfield  {author} {\bibinfo {author} {\bibfnamefont {I.}~\bibnamefont
  {Raychowdhury}},\ }\emph {\bibinfo {title} {{Prepotential Formulation of
  Lattice Gauge Theories}}},\ \href@noop {} {Ph.D. thesis},\ \bibinfo  {school}
  {Calcutta U.} (\bibinfo {year} {2013})\BibitemShut {NoStop}%
\bibitem [{\citenamefont {Zohar}\ and\ \citenamefont
  {Burrello}(2015)}]{Zohar2015}%
  \BibitemOpen
  \bibfield  {author} {\bibinfo {author} {\bibfnamefont {E.}~\bibnamefont
  {Zohar}}\ and\ \bibinfo {author} {\bibfnamefont {M.}~\bibnamefont
  {Burrello}},\ }\href {\doibase 10.1103/physrevd.91.054506} {\bibfield
  {journal} {\bibinfo  {journal} {Physical Review D}\ }\textbf {\bibinfo
  {volume} {91}} (\bibinfo {year} {2015}),\
  10.1103/physrevd.91.054506}\BibitemShut {NoStop}%
\bibitem [{\citenamefont {Zohar}\ \emph {et~al.}(2017)\citenamefont {Zohar},
  \citenamefont {Farace}, \citenamefont {Reznik},\ and\ \citenamefont
  {Cirac}}]{Zohar2017}%
  \BibitemOpen
  \bibfield  {author} {\bibinfo {author} {\bibfnamefont {E.}~\bibnamefont
  {Zohar}}, \bibinfo {author} {\bibfnamefont {A.}~\bibnamefont {Farace}},
  \bibinfo {author} {\bibfnamefont {B.}~\bibnamefont {Reznik}}, \ and\ \bibinfo
  {author} {\bibfnamefont {J.~I.}\ \bibnamefont {Cirac}},\ }\href {\doibase
  10.1103/physreva.95.023604} {\bibfield  {journal} {\bibinfo  {journal}
  {Physical Review A}\ }\textbf {\bibinfo {volume} {95}} (\bibinfo {year}
  {2017}),\ 10.1103/physreva.95.023604}\BibitemShut {NoStop}%
\bibitem [{\citenamefont {Alexandru}\ \emph {et~al.}(2019)\citenamefont
  {Alexandru}, \citenamefont {Bedaque}, \citenamefont {Harmalkar},
  \citenamefont {Lamm}, \citenamefont {Lawrence},\ and\ \citenamefont
  {Warrington}}]{Alexandru:2019nsa}%
  \BibitemOpen
  \bibfield  {author} {\bibinfo {author} {\bibfnamefont {A.}~\bibnamefont
  {Alexandru}}, \bibinfo {author} {\bibfnamefont {P.~F.}\ \bibnamefont
  {Bedaque}}, \bibinfo {author} {\bibfnamefont {S.}~\bibnamefont {Harmalkar}},
  \bibinfo {author} {\bibfnamefont {H.}~\bibnamefont {Lamm}}, \bibinfo {author}
  {\bibfnamefont {S.}~\bibnamefont {Lawrence}}, \ and\ \bibinfo {author}
  {\bibfnamefont {N.~C.}\ \bibnamefont {Warrington}} (\bibinfo {collaboration}
  {NuQS}),\ }\href {\doibase 10.1103/PhysRevD.100.114501} {\bibfield  {journal}
  {\bibinfo  {journal} {Phys. Rev. D}\ }\textbf {\bibinfo {volume} {100}},\
  \bibinfo {pages} {114501} (\bibinfo {year} {2019})},\ \Eprint
  {http://arxiv.org/abs/1906.11213} {arXiv:1906.11213 [hep-lat]} \BibitemShut
  {NoStop}%
\bibitem [{\citenamefont {Ji}\ \emph {et~al.}(2020)\citenamefont {Ji},
  \citenamefont {Lamm},\ and\ \citenamefont {Zhu}}]{Ji2020}%
  \BibitemOpen
  \bibfield  {author} {\bibinfo {author} {\bibfnamefont {Y.}~\bibnamefont
  {Ji}}, \bibinfo {author} {\bibfnamefont {H.}~\bibnamefont {Lamm}}, \ and\
  \bibinfo {author} {\bibfnamefont {S.}~\bibnamefont {Zhu}},\ }\href {\doibase
  10.1103/physrevd.102.114513} {\bibfield  {journal} {\bibinfo  {journal}
  {Physical Review D}\ }\textbf {\bibinfo {volume} {102}} (\bibinfo {year}
  {2020}),\ 10.1103/physrevd.102.114513}\BibitemShut {NoStop}%
\bibitem [{\citenamefont {Hackett}\ \emph {et~al.}(2019)\citenamefont
  {Hackett}, \citenamefont {Howe}, \citenamefont {Hughes}, \citenamefont {Jay},
  \citenamefont {Neil},\ and\ \citenamefont {Simone}}]{Hackett2019}%
  \BibitemOpen
  \bibfield  {author} {\bibinfo {author} {\bibfnamefont {D.~C.}\ \bibnamefont
  {Hackett}}, \bibinfo {author} {\bibfnamefont {K.}~\bibnamefont {Howe}},
  \bibinfo {author} {\bibfnamefont {C.}~\bibnamefont {Hughes}}, \bibinfo
  {author} {\bibfnamefont {W.}~\bibnamefont {Jay}}, \bibinfo {author}
  {\bibfnamefont {E.~T.}\ \bibnamefont {Neil}}, \ and\ \bibinfo {author}
  {\bibfnamefont {J.~N.}\ \bibnamefont {Simone}},\ }\href {\doibase
  10.1103/physreva.99.062341} {\bibfield  {journal} {\bibinfo  {journal}
  {Physical Review A}\ }\textbf {\bibinfo {volume} {99}} (\bibinfo {year}
  {2019}),\ 10.1103/physreva.99.062341}\BibitemShut {NoStop}%
\bibitem [{\citenamefont {Kreshchuk}\ \emph {et~al.}(2020)\citenamefont
  {Kreshchuk}, \citenamefont {Kirby}, \citenamefont {Goldstein}, \citenamefont
  {Beauchemin},\ and\ \citenamefont {Love}}]{kreshchuk2020quantum}%
  \BibitemOpen
  \bibfield  {author} {\bibinfo {author} {\bibfnamefont {M.}~\bibnamefont
  {Kreshchuk}}, \bibinfo {author} {\bibfnamefont {W.~M.}\ \bibnamefont
  {Kirby}}, \bibinfo {author} {\bibfnamefont {G.}~\bibnamefont {Goldstein}},
  \bibinfo {author} {\bibfnamefont {H.}~\bibnamefont {Beauchemin}}, \ and\
  \bibinfo {author} {\bibfnamefont {P.~J.}\ \bibnamefont {Love}},\ }\href@noop
  {} {\enquote {\bibinfo {title} {Quantum simulation of quantum field theory in
  the light-front formulation},}\ } (\bibinfo {year} {2020}),\ \Eprint
  {http://arxiv.org/abs/2002.04016} {arXiv:2002.04016 [quant-ph]} \BibitemShut
  {NoStop}%
\bibitem [{\citenamefont {Kreshchuk}\ \emph {et~al.}(2021)\citenamefont
  {Kreshchuk}, \citenamefont {Jia}, \citenamefont {Kirby}, \citenamefont
  {Goldstein}, \citenamefont {Vary},\ and\ \citenamefont
  {Love}}]{Kreshchuk2021}%
  \BibitemOpen
  \bibfield  {author} {\bibinfo {author} {\bibfnamefont {M.}~\bibnamefont
  {Kreshchuk}}, \bibinfo {author} {\bibfnamefont {S.}~\bibnamefont {Jia}},
  \bibinfo {author} {\bibfnamefont {W.~M.}\ \bibnamefont {Kirby}}, \bibinfo
  {author} {\bibfnamefont {G.}~\bibnamefont {Goldstein}}, \bibinfo {author}
  {\bibfnamefont {J.~P.}\ \bibnamefont {Vary}}, \ and\ \bibinfo {author}
  {\bibfnamefont {P.~J.}\ \bibnamefont {Love}},\ }\href {\doibase
  10.3390/e23050597} {\bibfield  {journal} {\bibinfo  {journal} {Entropy}\
  }\textbf {\bibinfo {volume} {23}},\ \bibinfo {pages} {597} (\bibinfo {year}
  {2021})}\BibitemShut {NoStop}%
\bibitem [{\citenamefont {Buser}\ \emph {et~al.}(2021)\citenamefont {Buser},
  \citenamefont {Gharibyan}, \citenamefont {Hanada}, \citenamefont {Honda},\
  and\ \citenamefont {Liu}}]{Buser:2020cvn}%
  \BibitemOpen
  \bibfield  {author} {\bibinfo {author} {\bibfnamefont {A.~J.}\ \bibnamefont
  {Buser}}, \bibinfo {author} {\bibfnamefont {H.}~\bibnamefont {Gharibyan}},
  \bibinfo {author} {\bibfnamefont {M.}~\bibnamefont {Hanada}}, \bibinfo
  {author} {\bibfnamefont {M.}~\bibnamefont {Honda}}, \ and\ \bibinfo {author}
  {\bibfnamefont {J.}~\bibnamefont {Liu}},\ }\href {\doibase
  10.1007/JHEP09(2021)034} {\bibfield  {journal} {\bibinfo  {journal} {JHEP}\
  }\textbf {\bibinfo {volume} {09}},\ \bibinfo {pages} {034} (\bibinfo {year}
  {2021})},\ \Eprint {http://arxiv.org/abs/2011.06576} {arXiv:2011.06576
  [hep-th]} \BibitemShut {NoStop}%
\bibitem [{\citenamefont {Kaplan}\ \emph {et~al.}(2003)\citenamefont {Kaplan},
  \citenamefont {Katz},\ and\ \citenamefont {Unsal}}]{Kaplan:2002wv}%
  \BibitemOpen
  \bibfield  {author} {\bibinfo {author} {\bibfnamefont {D.~B.}\ \bibnamefont
  {Kaplan}}, \bibinfo {author} {\bibfnamefont {E.}~\bibnamefont {Katz}}, \ and\
  \bibinfo {author} {\bibfnamefont {M.}~\bibnamefont {Unsal}},\ }\href
  {\doibase 10.1088/1126-6708/2003/05/037} {\bibfield  {journal} {\bibinfo
  {journal} {JHEP}\ }\textbf {\bibinfo {volume} {05}},\ \bibinfo {pages} {037}
  (\bibinfo {year} {2003})},\ \Eprint {http://arxiv.org/abs/hep-lat/0206019}
  {arXiv:hep-lat/0206019} \BibitemShut {NoStop}%
\bibitem [{\citenamefont {Carena}\ \emph {et~al.}(2021)\citenamefont {Carena},
  \citenamefont {Lamm}, \citenamefont {Li},\ and\ \citenamefont
  {Liu}}]{PhysRevD.104.094519}%
  \BibitemOpen
  \bibfield  {author} {\bibinfo {author} {\bibfnamefont {M.}~\bibnamefont
  {Carena}}, \bibinfo {author} {\bibfnamefont {H.}~\bibnamefont {Lamm}},
  \bibinfo {author} {\bibfnamefont {Y.-Y.}\ \bibnamefont {Li}}, \ and\ \bibinfo
  {author} {\bibfnamefont {W.}~\bibnamefont {Liu}},\ }\href {\doibase
  10.1103/PhysRevD.104.094519} {\bibfield  {journal} {\bibinfo  {journal}
  {Phys. Rev. D}\ }\textbf {\bibinfo {volume} {104}},\ \bibinfo {pages}
  {094519} (\bibinfo {year} {2021})}\BibitemShut {NoStop}%
\bibitem [{\citenamefont {Clemente}\ \emph {et~al.}(2022)\citenamefont
  {Clemente}, \citenamefont {Crippa},\ and\ \citenamefont
  {Jansen}}]{Clemente:2022cka}%
  \BibitemOpen
  \bibfield  {author} {\bibinfo {author} {\bibfnamefont {G.}~\bibnamefont
  {Clemente}}, \bibinfo {author} {\bibfnamefont {A.}~\bibnamefont {Crippa}}, \
  and\ \bibinfo {author} {\bibfnamefont {K.}~\bibnamefont {Jansen}},\
  }\href@noop {} {\  (\bibinfo {year} {2022})},\ \Eprint
  {http://arxiv.org/abs/2206.12454} {arXiv:2206.12454 [hep-lat]} \BibitemShut
  {NoStop}%
\bibitem [{\citenamefont {Carena}\ \emph {et~al.}(2022)\citenamefont {Carena},
  \citenamefont {Gustafson}, \citenamefont {Lamm}, \citenamefont {Li},\ and\
  \citenamefont {Liu}}]{Carena:2022hpz}%
  \BibitemOpen
  \bibfield  {author} {\bibinfo {author} {\bibfnamefont {M.}~\bibnamefont
  {Carena}}, \bibinfo {author} {\bibfnamefont {E.~J.}\ \bibnamefont
  {Gustafson}}, \bibinfo {author} {\bibfnamefont {H.}~\bibnamefont {Lamm}},
  \bibinfo {author} {\bibfnamefont {Y.-Y.}\ \bibnamefont {Li}}, \ and\ \bibinfo
  {author} {\bibfnamefont {W.}~\bibnamefont {Liu}},\ }\href@noop {} {\
  (\bibinfo {year} {2022})},\ \Eprint {http://arxiv.org/abs/2208.10417}
  {arXiv:2208.10417 [hep-lat]} \BibitemShut {NoStop}%
\bibitem [{\citenamefont {Preskill}(2018)}]{Preskill2018quantumcomputingin}%
  \BibitemOpen
  \bibfield  {author} {\bibinfo {author} {\bibfnamefont {J.}~\bibnamefont
  {Preskill}},\ }\href {\doibase 10.22331/q-2018-08-06-79} {\bibfield
  {journal} {\bibinfo  {journal} {{Quantum}}\ }\textbf {\bibinfo {volume}
  {2}},\ \bibinfo {pages} {79} (\bibinfo {year} {2018})}\BibitemShut {NoStop}%
\end{thebibliography}%

\end{document}